\pdfoutput=1
\documentclass[onecolumn, twoside, 11pt, draftclsnofoot]{IEEEtran}
\usepackage[T1]{fontenc}
\usepackage{stfloats, balance, enumerate, cite, xcolor, amsmath, mathtools, cuted, xcolor, mathrsfs, mathtools, amssymb}
\newcommand\myeqa{\mathrel{\overset{\makebox[0pt]{\mbox{\normalfont\tiny\sffamily (a)}}}{=}}}
\newcommand\myeqb{\mathrel{\overset{\makebox[0pt]{\mbox{\normalfont\tiny\sffamily (b)}}}{=}}}
\newcommand\myeqc{\mathrel{\overset{\makebox[0pt]{\mbox{\normalfont\tiny\sffamily (c)}}}{=}}}

\DeclareMathOperator*{\argmin}{\arg\!\min}
\DeclareMathOperator*{\argmax}{\arg\!\max}
\ifCLASSINFOpdf
   \usepackage[pdftex]{graphicx}
\else
\fi
\begin{document}

\title{User-Antenna Selection for Physical-Layer Network Coding based on Euclidean Distance}
\author{Vaibhav~Kumar,~\IEEEmembership{Student Member,~IEEE,} Barry~Cardiff,~\IEEEmembership{Member,~IEEE,} \\and~Mark~F.~Flanagan,~\IEEEmembership{Senior Member,~IEEE}
\thanks{The authors are with School of Electrical and Electronic Engineering, University College Dublin, Ireland (e-mail: vaibhav.kumar@ucdconnect.ie, \{barry.cardiff, mark.flanagan\}@ucd.ie).}
\thanks{Part of the content of this paper appeared in the Proc. of the IEEE Global Communications Conference (GLOBECOM'17), Singapore, 4-8 Dec. 2017. Part of the content of this paper also appeared in the Proc. of the IEEE International Symposium on Personal, Indoor and Mobile Radio Communications (PIMRC'17), Montreal, Canada, 8-13 Oct. 2017.}
\thanks{This publication has emanated from research conducted with the financial support of Science Foundation Ireland (SFI) and is co-funded under the European Regional Development Fund under Grant Number 13/RC/2077.}}

\markboth{Accepted in IEEE Transactions on Communications}%
{Kumar \MakeLowercase{\textit{et al.}}: User-Antenna Selection for PNC based on Euclidean Distance}

\let\bs\boldsymbol
\maketitle
\begin{abstract}
In this paper, we present the error performance analysis of a multiple-input multiple-output (MIMO) physical-layer network coding (PNC) system with two different user-antenna selection (AS) schemes in asymmetric channel conditions. For the first antenna selection scheme (AS1), where the user-antenna is selected in order to maximize the overall channel gain between the user and the relay, we give an explicit analytical proof that for binary modulations, the system achieves full diversity order of $\min(N_{A}, N_{B}) \times N_{R}$ in the multiple-access (MA) phase, where $N_{A}, N_{B}$ and $N_{R}$ denote the number of antennas at user $A$, user $B$ and relay $R$ respectively. We present a detailed investigation of the diversity order for the MIMO-PNC system with AS1 in the MA phase for any modulation order. A tight closed-form upper bound on the average SER is also derived for the special case when $N_{R} = 1$, which is valid for any modulation order. We show that in this case the system fails to achieve transmit diversity in the MA phase, as the system diversity order drops to 1 irrespective of the number of transmit antennas at the user nodes. Additionally, we propose a Euclidean distance (ED) based user-antenna selection scheme (AS2) which outperforms the first scheme in terms of error performance. Moreover, by deriving upper and lower bounds on the diversity order for the MIMO-PNC system with AS2, we show that this system enjoys both transmit and receive diversity, achieving full diversity order of $\min(N_A, N_B) \times N_R$ in the MA phase for any modulation order. Monte Carlo simulations are provided which confirm the correctness of the derived analytical results.
\end{abstract}


\IEEEpeerreviewmaketitle

\section{Introduction}
\IEEEPARstart{W}{ireless} PNC has received a lot of attention among researchers in recent years due to its inherent desirable properties of delay reduction, throughput enhancement and better spectral efficiency. The advantage of PNC can easily be seen in a two-way relay channel (TWRC), where bidirectional information exchange takes place in the half-duplex mode between two users $A$ and $B$ with the help of a relay $R$. In a TWRC, PNC requires only two time slots to exchange the information between the users compared to three time slots required by traditional network coding \cite{Liew}. In the first time slot, also termed the \emph{multiple access} (MA) phase, both users $A$ and $B$ simultaneously transmit their data to the relay $R$. Based on its received signal, the relay forms the maximum-likelihood (ML) estimate of the \emph{pair} of transmitted user constellation symbols. This estimate of the pair of user symbols is then mapped to a \emph{network-coded} constellation symbol using the denoise-and-forward (DNF) protocol \cite{Popovski} and the relay broadcasts this to both users in the next time slot, called the \emph{broadcast} (BC) phase. User constellation symbol pairs which are mapped to the same complex number in the network-coded constellation are said to form a \emph{cluster}. Using its own message transmitted in the previous MA phase, $A$ can decode the message transmitted from $B$ and vice versa. 

Different aspects of PNC relating to communication theory, information theory, wireless networking, finite-field and infinite-field PNC, as well as synchronization issues and the use of PNC for passive optical networks were discussed in \cite{Liew, Primer, FieldPNC}. The first software radio based implementation of PNC was reported in \cite{Katti}, together with a discussion on related problems and solutions.

A performance comparison among four time slot transmission scheme (non network-coded scheme), three time slot transmission scheme (network coding scheme) and two time slot transmission scheme (PNC scheme) for TWRCs in terms of bit-error rate (BER) and maximum sum-rate was presented in \cite{Branka_Globecom}. Closed-form expressions for a tight upper and lower bound on the average SER at the relay and tight bounds on the average end-to-end BER for a PNC system in a Rayleigh fading channel were presented in~\cite{MinKim}. An exact BER expression for a PNC system operating over a TWRC exhibiting fading was presented in \cite{PNC_BER} using Craig's polar coordinate form. A general framework for the symbol-error-rate (SER) performance analysis of PNC systems operating over AWGN channels was presented in \cite{SER}. In~\cite{Andrew}, a high signal-to-noise ratio (SNR) analysis for the error performance at the relay in a PNC system with binary or higher-order real/complex modulation as well as for real and complex channel coefficients was presented.

In~\cite{LinearVectorPNC}, a linear vector PNC scheme was proposed for an open-loop spatial MIMO TWRC, where no CSI was available at the users' end. An explicit solution for the network coding (NC) generator matrix to minimize the error probability at high SNR was proposed, and a novel closed-form expression for the average SER in Rayleigh fading was also presented. In~\cite{MIMO_LinearPNC}, a multiuser communication scenario was considered, where $K$ users simultaneously communicate with a receiver using space-time coded MIMO with linear PNC. All the user messages were encoded by the same linear dispersion space-time code and a novel iterative search algorithm was used to optimize the space-time coded linear PNC mapping. It was shown that the system achieves full rate and full diversity while achieving the maximum coding gain. 
However, despite the advantages of the linear MIMO-PNC systems proposed in~\cite{LinearVectorPNC} and~\cite{MIMO_LinearPNC}, the implementation cost of these systems are high -- all of the antennas from all users are utilized simultaneously to transmit the data, which require a large number of radio-frequency (RF) chains. One of the key differences between the MIMO-PNC systems proposed in~\cite{LinearVectorPNC},~\cite{MIMO_LinearPNC} and the MIMO-PNC
system proposed in this paper is that we take advantage of switched diversity (by virtue of user-antenna selection) to reduce the required number of RF chains - only one antenna per user is active during transmission. This reduces the overall cost of the system, while maintaining the full diversity order.

In the case of a \emph{fixed network coding} (FNC) system, the network code applied at the relay is always fixed and does not depend on channel conditions. One of the bottlenecks in the FNC system limiting the error performance is the existence of \emph{singular fade states} \cite{RajanPNC} that result in the phenomenon of distance shortening, which will be explained later in this paper. To solve this problem, a number of \emph{adaptive physical-layer network coding} (ANC) schemes have been proposed \cite{Akino, RajanPNC, AdaptiveOFDM}, where the relay adaptively selects the network mapping that offers the best performance based on the channel conditions. A similar scheme for multiple-input multiple-output (MIMO) two-way relaying was applied in~\cite{RajanMIMO}, and it was shown that the minimum distance between the network-coded constellation points at the relay becomes zero when all the rows of the channel matrix belong to a finite number of subspaces referred to as \emph{singular fade subspaces}. A computationally efficient analytical framework to choose the appropriate adaptive network codes at the relay for heterogeneous symmetric PNC was presented in \cite{HePNC_Rajan}.  A detailed introduction to wireless multi-way relaying using ANC was presented in \cite{MultiWay}. It was shown in \cite{LatinSquares} that every valid network mapping can be represented by a Latin square and that this relationship can be used to obtain network maps with optimized intercluster distance profiles. It has been shown in \cite{Akino, RajanPNC} that for a 4-ary modulation scheme in the MA phase, ANC may result in a 5-ary network map, and therefore a non-standard 5-ary modulation scheme will be required for the BC phase under certain channel conditions. Although ANC alleviates the problem of distance shortening in an efficient way, the related system complexity increases significantly due to the required clustering algorithm, and the increased cardinality of the relay's transmit constellation may incur a sacrifice in the reliability in the BC phase. 

In~\cite{Huang}, tight upper and lower bounds on the average BER of a multiple-antenna PNC system were presented, where the AS scheme based on the maximization of the overall channel gain between the user and the relay (which we will refer to as AS1 in the rest of this paper) is applied at both user nodes, and the users employ BPSK modulation. It was stated that for BPSK modulation, the MIMO-PNC system achieves a diversity order of $\min(N_{A}, N_{B}) \times N_{R}$ in the MA phase, and an explicit proof of the diversity order was provided for the special cases of the MISO-PNC system ($N_{A}, N_{B} > 1, N_{R} = 1$) and the SIMO-PNC system ($N_{A}, N_{B} = 1, N_{R} > 1$).

A popular paradigm for antenna selection in the literature is that based on the ED criterion, where the antenna at the transmitter node is selected such that the minimum ED between different symbols in the received constellation is maximized. Such an antenna selection scheme was discussed in~\cite{Paulraj} for spatial multiplexing (SMx) systems, in~\cite{PNC_Scheduling} for opportunistic PNC scheduling and in~\cite{Hari} for spatial modulation (SM). In~\cite{Paulraj}, a SMx system with $\mathscr{N}_t$ transmit antennas, $\mathscr{N}_r$ receive antennas, and a $1~:~\mathscr N (\mathscr N_t > \mathscr N, \mathscr N_r>\mathscr N)$ multiplexer was considered, where a low-bandwidth, zero-delay, error-free feedback path indicated the optimal $\mathscr N$ of $\mathscr N_t$ antennas for transmission, computed using current channel state information at the receiver. For the ML receiver, the authors proposed to choose the subset of transmit antennas which resulted in a constellation with largest minimum ED. It was shown that this ED based AS resulted in minimum error rate for the SMx system based on the ML receiver. In~\cite{PNC_Scheduling}, a three-way wireless communication system was considered, where each user desires to transmit independent data to other users via a relay. Since the overall throughput of such a system is limited by the worst channel, a scheduling system employing PNC was considered to optimize the overall system throughput. It was shown that the selection of a user-pair based on the largest minimum ED between the superposed constellation at the relay resulted into better overall throughput compared to that of the channel-norm based user-pair selection and round-robin based scheduling. In~\cite{Hari}, a comprehensive analysis of the transmit diversity order for the ED based antenna selection scheme in an SM system (consisting of $\mathscr N_t$ transmit antennas and $\mathscr N_r$ receive antennas) was presented. For each transmission, $\mathscr N_{SM}$ out of $\mathscr N_t$ transmit antennas were selected in order to achieve spatial switching gain (SSG). It was proved explicitly that such an SM system enjoys both transmit and receive diversity, achieving a diversity order of $\mathscr N_r (\mathscr N_t - \mathscr N_{SM} + 1)$.

It is important to note that the error performance analysis in the BC phase of the PNC system in a TWRC is similar to a traditional (non network-coded) point-to-point communication system and the end-to-end error performance of the PNC system will be dominated by the error performance in the MA phase. Therefore, in this paper we analyze the error performance and the diversity order of the MIMO-PNC system in the MA phase only. The main contributions of this paper are summarized as follows:
\begin{itemize}
	\item We give an explicit analytical proof that the diversity order of the MIMO-PNC system with binary modulation and AS1 is equal to $\min(N_{A}, N_{B}) \times N_{R}$.
	\item We provide a detailed investigation of the error rate performance and diversity order of the MIMO-PNC system with AS1 for any modulation order $M$. A closed-form expression for a tight upper bound on the average SER is derived for the special case when $N_{R} = 1$. The presented diversity analysis confirms that the performance of the MIMO-PNC system with AS1 degrades severely for non-binary modulations due to the distance shortening phenomenon at the relay, and the system fails to achieve transmit diversity. We give an analytical proof that the diversity order of such system drops to 1 for the case when $N_{R} = 1$.
	\item We propose an ED based AS scheme (which we will refer to as AS2 in the rest of this paper) for the MIMO-PNC system to mitigate the deleterious effects of distance shortening at the relay. Furthermore, we derive upper and lower bounds on the diversity order and prove that the system with AS2 achieves a full diversity order of $\min(N_{A}, N_{B})~\times~N_{R}$ for any modulation order.
\end{itemize}

The rest of this paper is organized as follows: In Section II, we present the system model for the MIMO-PNC system. In Section III, we introduce the AS1 and AS2 schemes and give an illustration of the performance superiority of AS2 over AS1. In Section IV, we present a comprehensive error performance analysis of the MIMO-PNC system with AS1 and also provide a diversity analysis. Section V deals with the derivation of upper and lower bounds on the diversity order of the MIMO-PNC system with AS2. In Section VI, we present extensive simulation and analytical results along with discussion. Finally, the conclusion is presented in Section VII.

\section{System Model}
\begin{figure}[t]
\centering
\includegraphics[scale=0.9859]{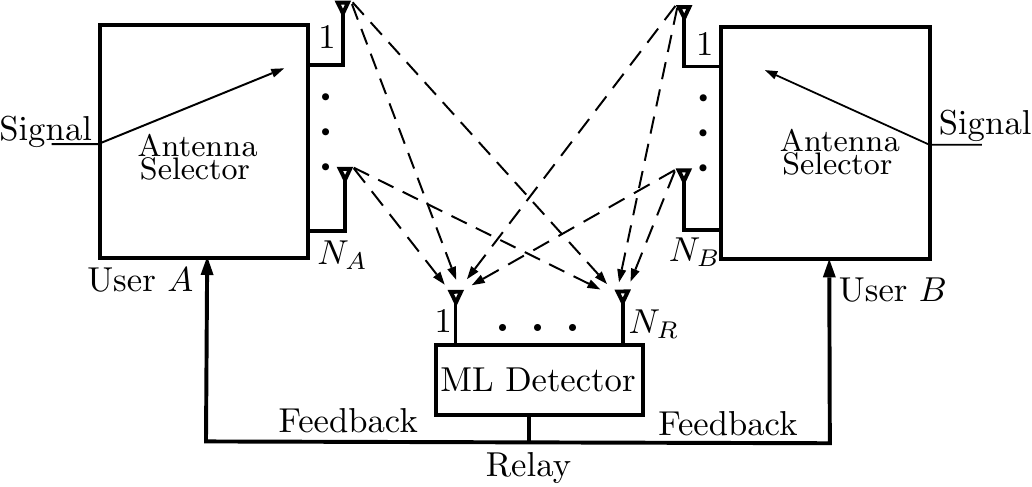}
\caption{System model for PNC system with multiple antennas at user and relay nodes.}
\label{SysMod}
\end{figure}

The system model for the MIMO-PNC system is shown in Fig.~\ref{SysMod} where two users $A$ and $B$ are equipped with $N_{A}~>~1$ and $N_{B}~>~1$ antennas, respectively, while relay $R$ is equipped with $N_{R} \geq 1$ antennas. 
During the MA phase, only one of the antennas from each user is used for signal transmission, and the choice of antennas is based on feedback received from the relay. The channel between user $m \in \{A, B\}$ and the relay $R$ is modeled as slow Rayleigh fading with perfect CSI available at $R$ only. We assume that the channel remains constant during a frame transmission and changes independently from one frame to another. Hence the channel coefficient between the $i^{\text{th}}$ antenna of user $m$ and the relay is distributed according to $\mathcal{CN}(0, 1)$. Both users employ the same unit-energy $M$-ary constellation $\mathcal{X}$, and $\Delta \mathcal{X}$ denotes the difference constellation set of $\mathcal{X}$, defined as $\Delta \mathcal{X} \triangleq \{\Delta x = x - x^{\prime}|x, x^{\prime} \in \mathcal{X}\}$. 

Let $s_{m} \in \mathbb{Z}_{M} = \{0, 1, \ldots, M-1\}$ denote the message symbol at user $m$, and $x_{m} = \mathcal{F}(s_{m}) \in \mathcal{X}$ denote the corresponding transmitted constellation symbol, where $\mathcal{F}$ denotes the constellation mapping function. 
The signal vector received at the relay during the MA phase is
\begin{equation}
	\boldsymbol{y} = \sqrt{E_{A}}\,\boldsymbol{h}_{A} x_{A} + \sqrt{E_{B}}\, \boldsymbol{h}_{B} x_{B} + \boldsymbol{n}, \label{signal_received}
\end{equation}
where $\boldsymbol{n} \in \mathbb{C}^{N_{R} \times 1}$ denotes the noise vector at the relay whose elements are assumed to be distributed according to $\mathcal{CN}(0, N_{0})$, $E_{m}$ denotes the energy of the transmitted signal from user $m$ and $\boldsymbol{h}_{m}  = [h_{m, 1} \, h_{m, 2}\, \cdots\, h_{m, N_{R}}]^{T}\in \mathbb{C}^{N_{R} \times 1}$ is the channel coefficient vector of the link between the selected antenna of user $m$ and the relay antennas (based on the AS scheme). The relay's goal is to determine the network-coded symbol $s_R \triangleq \mathcal M_s(s_A, s_B)$, where $\mathcal M_s: \mathbb Z_M^2 \to \mathbb Z_M$ is the PNC mapping, or equivalently to determine the corresponding constellation symbol $x_R = \mathcal F (s_R) \triangleq \mathcal M_c(x_A, x_B)$, where $\mathcal M_c : \mathcal X^2 \to \mathcal X$ represents the ``constellation-domain'' version of the PNC mapping. Table~\ref{mapping_table} shows an example PNC mapping for QPSK modulation, where the PNC mapping $\mathcal M_s: \mathbb Z_M^2 \to \mathbb Z_M$ represents \emph{bitwise} addition (XOR) in $\mathbb Z_4$.

The relay will form an estimate of $x_R$, denoted by $\hat{x}_R$, as follows. First, the relay computes the ML estimate of the transmitted symbol pair $(x_{A}, x_{B}) \in \mathcal{X}^{2}$ given by
\begin{equation}
	\left(\hat{x}_{A}, \hat{x}_{B}\right) = \argmin_{(x_{A}, x_{B}) \in \mathcal{X}^{2}} \left\Vert \boldsymbol{y} - \sqrt{E_{A}}\, \boldsymbol{h}_{A}x_{A} - \sqrt{E_{B}}\, \boldsymbol{h}_{B}x_{B} \right\Vert. \label{MLE}
\end{equation}

\renewcommand{\tabcolsep}{4pt}
\begin{table}[t]
\centering
\caption{Example PNC mapping at the relay for QPSK constellation.}
\label{mapping_table}
\begin{tabular}{|c|c|c|c|}
\hline
$(x_{A}, x_{B})$                 &$(s_{A}, s_{B})$  & \begin{tabular}[c]{@{}c@{}}$s_{R}$\\ = $\mathcal{M}_{s}(s_{A}, s_{B})$\end{tabular}  & \begin{tabular}[c]{@{}c@{}}$x_{R}$\\ = $\mathcal{F}(s_{R})$\end{tabular}   \\ \hline
\begin{tabular}[c]{@{}c@{}}$\left(\frac{1+i}{\sqrt{2}}, \frac{1+i}{\sqrt{2}} \right)$, $\left(\frac{-1+i}{\sqrt{2}}, \frac{-1+i}{\sqrt{2}}\right)$,\\ $\left(\frac{-1-i}{\sqrt{2}}, \frac{-1-i}{\sqrt{2}}\right)$, $\left(\frac{1-i}{\sqrt{2}}, \frac{1-i}{\sqrt{2}}\right)$\end{tabular} & \begin{tabular}[c]{@{}c@{}}(0, 0), (1, 1),\\ (2, 2), (3, 3)\end{tabular}  & 0       & $\frac{1+i}{\sqrt{2}}$     \\ \hline
\begin{tabular}[c]{@{}c@{}}$\left(\frac{1+i}{\sqrt{2}}, \frac{-1+i}{\sqrt{2}}\right)$, $\left(\frac{-1+i}{\sqrt{2}}, \frac{1+i}{\sqrt{2}}\right)$,\\ $\left(\frac{-1-i}{\sqrt{2}}, \frac{1-i}{\sqrt{2}}\right)$, $\left(\frac{1-i}{\sqrt{2}}, \frac{-1-i}{\sqrt{2}}\right)$\end{tabular} & \begin{tabular}[c]{@{}c@{}}(0, 1), (1, 0),\\ (2, 3), (3, 2)\end{tabular} & 1       & $\frac{-1+i}{\sqrt{2}}$    \\ \hline
\begin{tabular}[c]{@{}c@{}}$\left(\frac{1+i}{\sqrt{2}}, \frac{-1-i}{\sqrt{2}}\right)$, $\left(\frac{-1-i}{\sqrt{2}}, \frac{1+i}{\sqrt{2}}\right)$,\\ $\left(\frac{-1+i}{\sqrt{2}}, \frac{1-i}{\sqrt{2}}\right)$, $\left(\frac{1-i}{\sqrt{2}}, \frac{-1+i}{\sqrt{2}}\right)$\end{tabular} & \begin{tabular}[c]{@{}c@{}}(0, 2), (2,0),\\ (1, 3), (3,1)\end{tabular}    & 2       & $\frac{-1-i}{\sqrt{2}}$    \\ \hline
\begin{tabular}[c]{@{}c@{}}$\left(\frac{1+i}{\sqrt{2}}, \frac{1-i}{\sqrt{2}}\right)$, $\left(\frac{1-i}{\sqrt{2}}, \frac{1+i}{\sqrt{2}}\right)$,\\ $\left(\frac{-1+i}{\sqrt{2}}, \frac{-1-i}{\sqrt{2}}\right)$, $\left(\frac{-1-i}{\sqrt{2}}, \frac{-1+i}{\sqrt{2}}\right)$\end{tabular} & \begin{tabular}[c]{@{}c@{}}(0, 3), (3,0),\\ (1, 2), (2, 1)\end{tabular}  & 3       & $\frac{1-i}{\sqrt{2}}$     \\ \hline
\end{tabular}
\end{table}
\noindent Having this joint estimate $(\hat{x}_{A}, \hat{x}_{B}) \in \mathcal{X}^{2}$, the relay calculates $\hat x_R = \mathcal M_c(\hat x_A, \hat x_B)$. The error performance at the relay will depend on the minimum distance between the signal points in different clusters, defined as
\begin{align}
d_{\min}(\boldsymbol{h}_{A}, \boldsymbol{h}_{B}) \triangleq \min_{\substack{ (x_{A}, x_{B}), (x_{A}', x_{B}') \in \mathcal{X}^{2} \\ \mathcal{M}_{c} (x_{A}, x_{B}\!) \neq \mathcal{M}_{ c}(x_{A}', x_{B}')}} \left\Vert \sqrt{E_{A}}\boldsymbol{h}_{A}(x_{A} - x_{A}') +\! \sqrt{E_{B}}\boldsymbol{h}_{B}(x_{B} - x_{B}')\right\Vert, \label{d_min}
\end{align} 
It is clear from \eqref{d_min} that the value of $d_{\min}$ depends on the channel between the users and the relay. In general, when the values of channel coefficients are such that the distance between the clusters is significantly reduced, the phenomenon is called \emph{distance shortening}. 

In the next section, we introduce two different AS schemes and explain the performance superiority of one over the other with the help of an example.

\section{User-Antenna Selection for PNC}
This section presents two different AS schemes for the MIMO-PNC system. In the first scheme (AS1), the user-antenna is selected in order to maximize the overall channel gain between the user and the relay. Hence we define
\begin{equation}
	\mathfrak{z}_{m} \triangleq \max_{1 \leq i \leq N_{m}} \sum_{j = 1}^{N_{R}} \vert \mathfrak{h}_{m, i, j} \vert^{2}, \label{index_AS1}
\end{equation}
where $\mathfrak{h}_{m, i, j} \sim \mathcal{CN}(0, 1)$ is the channel coefficient between the $i^{\text{th}}$ antenna of user $m$ and $j^{\text{th}}$ antenna of relay $R$. Since $h_{m, j}$ is the channel coefficient between the \emph{selected} antenna of user $m$ and $j^{\text{th}}$ antenna of $R$, we may write
\begin{equation}
	\mathfrak{z}_{m} = \sum_{j = 1}^{N_{R}} \left\vert h_{m, j}\right\vert^{2}. \label{z_def}
\end{equation}
We present the probability density function (PDF) of $\mathfrak{z}_{m}$ in the following proposition.
\paragraph*{Proposition 1} The PDF of $\mathfrak{z}_{m}$ is given by
\begin{align}
	f(\mathfrak{z}_{m}) & = \dfrac{N_{m}}{(N_{R} - 1)!} \sum_{ \substack{k_{0} + k_{1} + \cdots + k_{N_{R}} \\ = N_{m} - 1}} \binom{N_{m} - 1}{k_{0}, \ldots, k_{N_{R}}} (-1)^{N_{m} - 1 - k_{0}}  \left[ \prod_{j = 0}^{N_{R} - 1} \left( \dfrac{1}{j!} \right)^{k_{j + 1}} \right] \notag \\
	& \hspace{8cm} \times \mathfrak{z}_{m}^{N_{R} + s - 1} \exp [-(N_{m} - k_{0}) \mathfrak{z}_{m}].\label{fz}
\end{align}
\paragraph*{Proof} See Appendix A.
\begin{figure}[b]
\centering
\begin{minipage}{.45\textwidth}
  \centering
  \includegraphics[scale = 0.355]{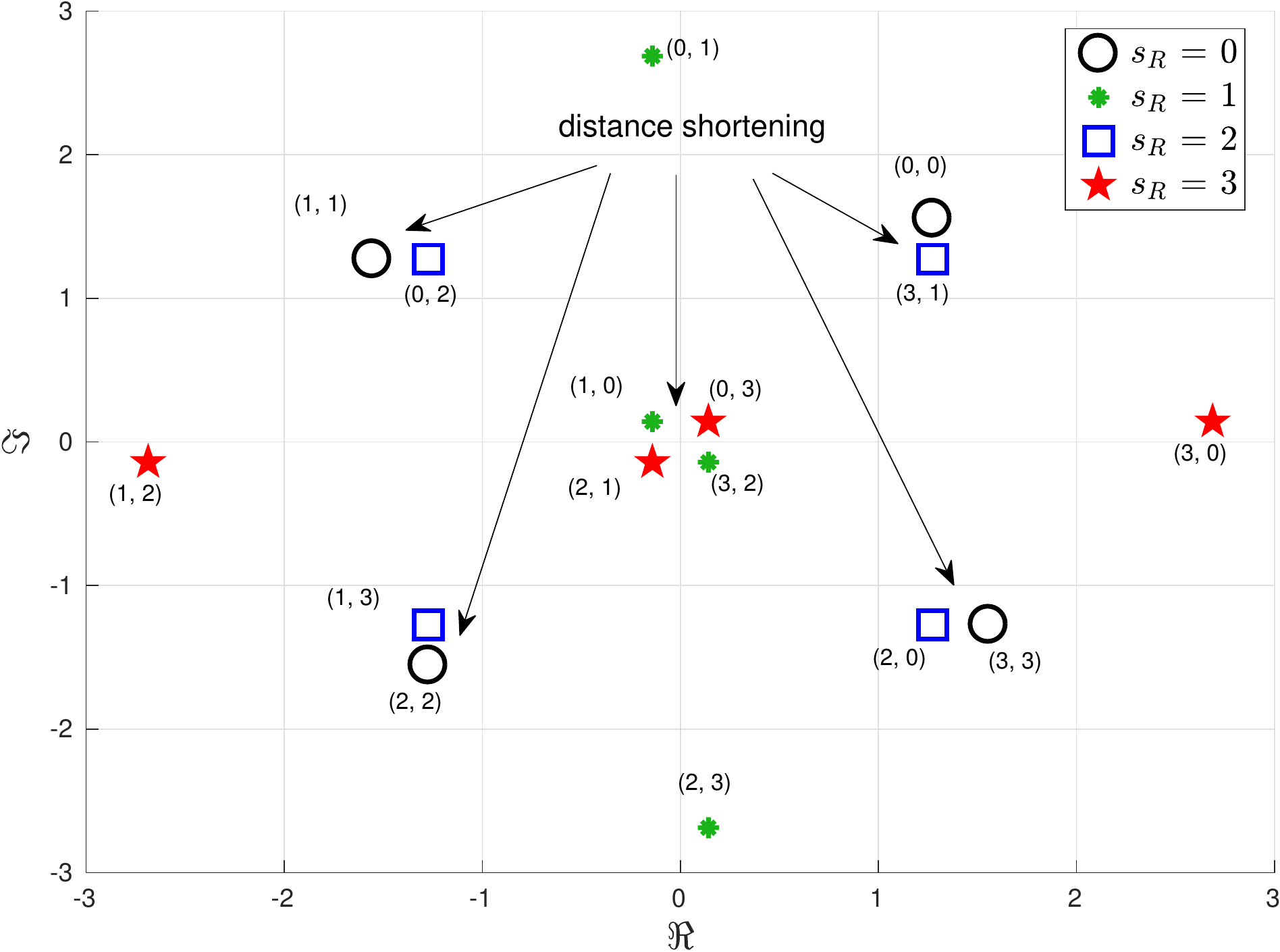}
  \caption{Noise-free received signal constellation ($h_{A} x_{A} + h_{B} x_{B}$) at the relay using AS1 with $N_{A} = N_{B} = 2, N_{R} = 1$ and $E_{A} = E_{B} = 1$. Markers of different color/shape close to each other show the distance shortening phenomenon.}
  \label{AS1_Const}
\end{minipage}%
\hfill
\begin{minipage}{.45\textwidth}
  \centering
  \vspace{-0.6cm}
  \includegraphics[scale = 0.48]{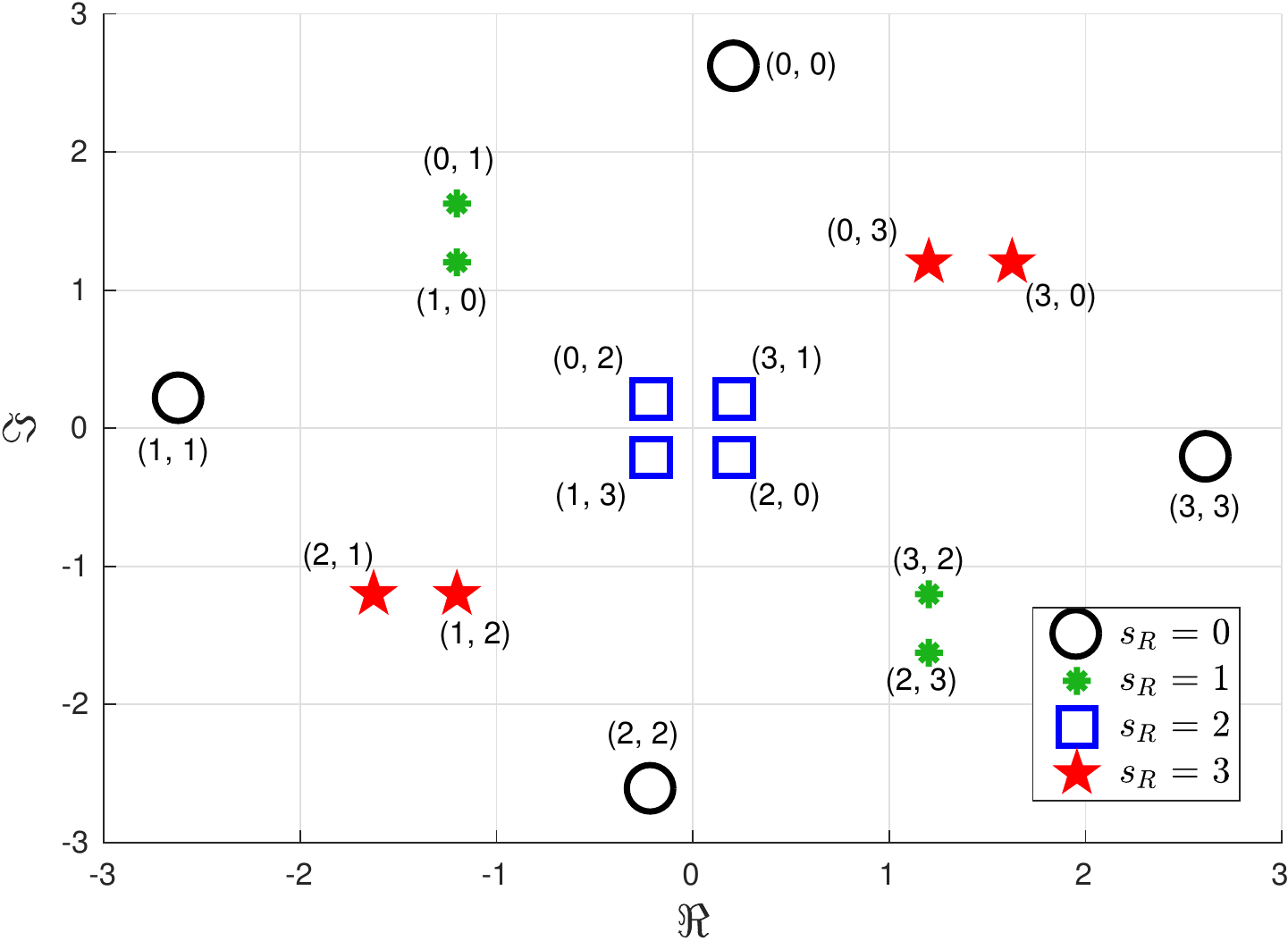}
  \caption{Noise-free received signal constellation ($h_{A} x_{A} + h_{B} x_{B}$) at the relay using AS2 with $N_{A} = N_{B} = 2, N_{R} = 1$ and $E_{A} = E_{B} = 1$. Markers of different color/shape are far apart from each other.}
  \label{AS2_Const}
\end{minipage}
\end{figure}

In contrast to this, in the second scheme (AS2) the user-antenna of each user is selected such that the minimum ED between the clusters at the relay is maximized. Let $\mathcal{I} = \{(i, j): 1 \leq i \leq N_{A}, 1 \leq j \leq N_{B}\}$ be the set which enumerates all of the possible $n = N_{A} \times N_{B}$ combinations of selecting one antenna from each user. Among these $n$ combinations, the set of user-antennas that maximizes the minimum ED between the clusters is obtained as
\begin{equation}
	I_{ED} \!=\! \argmax_{I \in \mathcal{I}} \left\{ \min_{\substack{\boldsymbol{x}, \boldsymbol{x}^{\prime} \in \mathcal{X}^{2} \\ \mathcal{M}_{c}(\boldsymbol{x}) \neq \mathcal{M}_{c}(\boldsymbol{x}^{\prime})}}\!\!\!\! \left\Vert \boldsymbol{H}_{I}\left(\begin{bmatrix}
	\sqrt{E_{A}}x_{A} \\ \sqrt{E_{B}} x_{B} \end{bmatrix}	 - \begin{bmatrix}
	\sqrt{E_{A}}x_{A}^{\prime} \\ \sqrt{E_{B}} x_{B}^{\prime} \end{bmatrix}\right)\right\Vert^{2}\right\}, \label{index_AS2}
\end{equation}
where $\boldsymbol{H}_{I} = [\boldsymbol{\mathfrak{h}}_{A, i} \ \boldsymbol{\mathfrak{h}}_{B, j}] \in \mathbb{C}^{N_{R} \times 2}$, $\boldsymbol{\mathfrak{h}}_{A, i} = [\mathfrak{h}_{A, i, 1} \ \ldots \ \mathfrak{h}_{A, i, N_{R}}]^{T}$, $\boldsymbol{\mathfrak{h}}_{B, j} = [\mathfrak{h}_{B, j, 1} \ \ldots \ \mathfrak{h}_{B, j, N_{R}}]^{T}$, and $\boldsymbol{H}_{I_{ED}} = [\boldsymbol{h}_{A} \ \boldsymbol{h}_{B}] \in \mathbb{C}^{N_{R} \times 2}$ is the optimal channel matrix. 

To understand the performance superiority of AS2 over AS1, we first consider a simple example of one transmission slot where the users transmit their messages using QPSK modulation. 

Suppose that $N_{A} = N_{B} = 2$, $N_{R} = 1$ and $\mathfrak{h}_{A, 1, 1} = (1+i)/\sqrt{2}$, $\mathfrak{h}_{A, 2, 1} = (1 - 0.5i)/\sqrt{2}$, $\mathfrak{h}_{B, 1, 1} = (1 - 0.8i)/\sqrt{2}$ and $\mathfrak{h}_{B, 2, 1} = (1+0.7i)/\sqrt{2}$. In this case, since $\left\vert\mathfrak{h}_{A, 1, 1}\right\vert > \left\vert \mathfrak h_{A, 2, 1} \right\vert$ and  $\left\vert\mathfrak{h}_{B, 1, 1}\right\vert > \left\vert \mathfrak h_{B, 2, 1} \right\vert$, AS1 will select the antenna combination $I = (1, 1)$. With this combination the minimum distance between the clusters at the relay becomes very small, which can lead to an incorrect ML estimate at the relay. Fig.~\ref{AS1_Const} shows a plot, for AS1, of the noise-free received signal at the relay, i.e., $h_{A}x_{A} + h_{B}x_{B}$ (here we assume $E_{A} = E_{B} = 1$), together with the corresponding network-coded symbols, where each $2$-tuple in the figure represents $(s_{A}, s_{B})$.   

In contrast to this, the proposed antenna selection scheme (AS2) chooses $I = I_{ED} = (1, 2)$ as the optimal combination and the resulting network-coded symbols are shown in Fig.~\ref{AS2_Const}. It is clear that AS2 overcomes the distance shortening phenomenon. The following section presents the error performance analysis and the diversity analysis of the MIMO-PNC system with AS1.
\section{AS1: Antenna selection based on the maximum overall channel gain}
For the error performance analysis of AS1, we use the \emph{union-bound} approach given in \cite{RajanPNC} rather that the approach given in \cite{Huang} which applies only for binary modulations. Using~\cite[eq.~(6)]{RajanPNC}, the average SER for FNC is given by
\begin{align}
	P_{e} = & \dfrac{1}{M^{2}} \sum_{(x_{A}, x_{B}) \in \mathcal{X}^{2}} \sum_{\substack{(x_{A}, x_{B}) \neq (x_{A}^{\prime}, x_{B}^{\prime}) \in \mathcal{X}^{2} \\ \mathcal{M}_{c}(x_{A}, x_{B}) \neq  \mathcal{M}_{c}(x_{A}^{\prime}, x_{B}^{\prime})}} \mathbb{E} \left[\mathcal{P}\left\{(x_{A}, x_{B}) \rightarrow (x_{A}^{\prime}, x_{B}^{\prime}) \mid \boldsymbol{h}_{A}, \boldsymbol{h}_{B}\right\} \right] \nonumber \\
	= & \dfrac{1}{M^{2}}\! \left[ \sum_{(x_{A}, x_{B}) \in \mathcal{X}^{2}} \sum_{x_{A} \neq x_{A}^{\prime} \in \mathcal{X}} \mathbb{E} \left[\mathcal{P} \{ (x_{A}, x_{B})\!\! \to \!\!(x_{A}^{\prime}, x_{B}) \!\mid \!\boldsymbol{h}_{A}, \boldsymbol{h}_{B} \} \right] \right. \notag \\
	&  \hspace{0.5cm}+ \sum_{(x_{A}, x_{B}) \in \mathcal{X}^{2}} \sum_{x_{B} \neq x_{B}^{\prime} \in \mathcal{X}} \mathbb{E} \left[\mathcal{P} \{ (x_{A}, x_{B})\!\! \to\!\! (x_{A}, x_{B}^{\prime})\! \mid\! \boldsymbol{h}_{A}, \boldsymbol{h}_{B} \}\right]  \nonumber \\
	&   \hspace{0.5cm}+ \sum_{(x_{A}, x_{B}) \in \mathcal{X}^{2}} \sum_{\substack{ x_{A} \neq x_{A}^{\prime} \in \mathcal{X} \\ x_{B} \neq x_{B}^{\prime} \in \mathcal{X}}} \mathbb{E} \left[\mathcal{P} \left\{ (x_{A}, x_{B}) \rightarrow (x_{A}^{\prime}, x_{B}^{\prime}),\mathcal{M}_{c}(x_{A}, x_{B}) \neq \mathcal{M}_{c}(x_{A}^{\prime}, x_{B}^{\prime}) \mid \boldsymbol{h}_{A}, \boldsymbol{h}_{B}\right\} \right]\Bigg], \label{Pe_exact}
\end{align}
where $\mathbb{E}[\cdot]$ is the expectation operator (here the expectation is performed with respect to $\boldsymbol{h}_{A}$ and $\boldsymbol{h}_{B}$) and $\mathcal{P}\{(\mathfrak{x}_{1},~\mathfrak{x}_{2})~\to~(\mathfrak{x}_{1}',~\mathfrak{x}_{2}')\}$ denotes the pairwise error probability, i.e., the probability that the signal pair $(\mathfrak{x}_{1}', \mathfrak{x}_{2}')$ is more likely than $(\mathfrak{x}_{1}, \mathfrak{x}_{2})$ from the receiver's perspective. An upper bound on the average SER can be given by
\begin{align}	
	P_{e} \leq \dfrac{1}{M^{2}} \sum_{(x_{A}, x_{B}) \in \mathcal{X}^{2}} & \sum_{ \substack{ (x_{A}, x_{B}) \neq (x_{A}^{\prime}, x_{B}^{\prime}) \in \mathcal{X}^{2} \\ \mathcal{M}_{c}(x_{A}, x_{B}) \neq \mathcal{M}_{c} (x_{A}^{\prime}, x_{B}^{\prime })} }\mathbb{E} \left[ Q \left( \dfrac{1}{\sqrt{2N_{0}}} \left\Vert \sum_{m \in \{A, B\}}\sqrt{E_{m}}\boldsymbol{h}_{m}(x_{m} - x_{m}') \right\Vert \right) \right], \notag
\end{align}
where $Q(\cdot)$ is the Gaussian Q-function. The Chernoff bound on the Q-function used in \cite{RajanPNC} results in a loose upper bound for the present case and hence we use the Chiani approximation~\cite[eq.~(14)]{Chiani} instead, yielding
\begin{align}
	P_{e} \lesssim & \dfrac{1}{M^{2}} \!\!\!\!\sum_{(x_{A}, x_{B}) \in \mathcal{X}^{2}} \!\!\sum_{\substack{(x_{A}, x_{B}) \neq (x_{A}^{\prime}, x_{B}^{\prime}) \in \mathcal{X}^{2} \\ \mathcal{M}_{c}(x_{A}, x_{B}) \neq  \mathcal{M}_{c}(x_{A}^{\prime}, x_{B}^{\prime})}} \!\!\!\!\left[\! \dfrac{1}{12} \mathbb{E} \left\{ \!\exp\! \left(\dfrac{-\left\Vert \sum_{m \in \{A, B\}}\sqrt{E_{m}}\boldsymbol{h}_{m} \Delta x_{m} \right\Vert^{2}}{4N_{0}}\right) \right\} \right. \notag \\
	& \hspace{6cm} \left. + \dfrac{1}{4} \mathbb{E} \left\{ \!\exp \!\left(\dfrac{-\left\Vert \sum_{m \in \{A, B\}}\sqrt{E_{m}}\boldsymbol{h}_{m} \Delta x_{m} \right\Vert^{2}}{3N_{0}}\right) \right\} \right] \nonumber \\
	 = & \dfrac{1}{M^{2}} \sum_{(x_{A}, x_{B}) \in \mathcal{X}^{2}} \sum_{\substack{(x_{A}, x_{B}) \neq (x_{A}^{\prime}, x_{B}^{\prime}) \in \mathcal{X}^{2} \\ \mathcal{M}_{c}(x_{A}, x_{B}) \neq  \mathcal{M}_{c}(x_{A}^{\prime}, x_{B}^{\prime})}}\left[\dfrac{\mathbb{E}(\Upsilon_{1})}{12}  + \dfrac{\mathbb{E}(\Upsilon_{2})}{4} \right]. \label{ChianiApp}
\end{align}

Now we analyze the three different terms on the right-hand side of \eqref{Pe_exact} separately as follows: 

\subsubsection*{Case I -- When $x_{A} \neq x_{A}'$ and $x_{B} = x_{B}'$}
In this case $\Delta x_{B} = 0$ and hence 
\begin{equation}
	\Upsilon_{1} \!\!=\! \exp \left(\!\dfrac{-E_{A}}{4N_{0}} \left\Vert \boldsymbol{h}_{A} \Delta x_{A} \right\Vert^{2} \!\right)\! =\! \exp \left(\!\!\dfrac{-E_{A}|\Delta x_{A}|^{2}}{4N_{0}}\! \sum_{j = 1}^{N_{R}} \left\vert h_{A, j} \right\vert^{2} \!\right),\!\! \label{Upsilon1_Case1}
\end{equation} 
and
\begin{equation}
	\Upsilon_{2}\!\! =\! \exp \left(\!\dfrac{-E_{A}}{3N_{0}} \left\Vert \boldsymbol{h}_{A} \Delta x_{A} \right\Vert^{2} \!\right)\! = \!\exp \left(\!\!\dfrac{-E_{A}|\Delta x_{A}|^{2}}{3N_{0}} \!\sum_{j = 1}^{N_{R}} \left\vert h_{A, j} \right\vert^{2} \!\right). \notag
\end{equation}

Defining $\Theta_{A, 1} \triangleq \mathbb{E}[\Upsilon_{1}]/12$ and $\Theta_{A, 2} \triangleq \mathbb{E}[\Upsilon_{2}]/4$, the average SER arising from the case when $x_{A} \neq x_{A}'$ and $x_{B} = x_{B}'$ can be written as
\begin{align}
	\mathcal{P} \left\{ (x_{A}, x_{B}) \to (x_{A}', x_{B})\right\}\lesssim \dfrac{1}{M^{2}} \sum_{(x_{A}, x_{B}) \in \mathcal{X}^{2}} \sum_{x_{A} \neq x_{A}' \in \mathcal{X}} \left( \Theta_{A, 1} + \Theta_{A, 2}\right), \label{error_case1}
\end{align}
where 
\begin{align}
	\Theta_{A, 1} = & \dfrac{N_{A}}{12(N_{R}-1)!} \sum_{\substack{ k_{0} + k_{1} + \cdots + k_{N_{R}} \\ = N_{A}-1}} \binom{N_{A} - 1}{k_{0}, \ldots, k_{N_{R}}} (-1)^{N_{A} -1 - k_{0}} \left[ \prod_{j = 0}^{N_{R} - 1} \left( \dfrac{1}{j!} \right)^{k_{j + 1}} \right] \notag \\
	& \hspace{5.5cm}\times \left( \dfrac{E_{A} |\Delta x_{A}|^{2}}{4N_{0}} + N_{A} - k_{0} \right)^{-(N_{R} + s)}(N_{R} + s - 1)!. \label{ThetaA1}
\end{align}
\begin{align}
	\Theta_{A, 2} = & \dfrac{N_{A}}{4(N_{R}-1)!} \sum_{\substack{ k_{0} + k_{1} + \cdots + k_{N_{R}} \\ = N_{A}-1}} \binom{N_{A} - 1}{k_{0}, \ldots, k_{N_{R}}} (-1)^{N_{A} -1 - k_{0}} \left[ \prod_{j = 0}^{N_{R} - 1} \left( \dfrac{1}{j!} \right)^{k_{j + 1}} \right] \notag \\
	& \hspace{5.5cm} \times \left( \dfrac{E_{A} |\Delta x_{A}|^{2}}{3N_{0}} + N_{A} - k_{0} \right)^{-(N_{R} + s)} (N_{R} + s - 1)!. \label{ThetaA2}
\end{align}
The derivation of the closed-form expression for $\Theta_{A, 1}$ is presented in~Appendix~B and the closed-form expression for $\Theta_{A, 2}$ can be derived in the same fashion.
\subsubsection*{Case II -- When $x_{A} = x_{A}'$ and $x_{B} \neq x_{B}'$}
In this case $\Delta x_{A} = 0$ and hence the average SER arising from the case when $x_{A}~=~x_{A}'$ and $x_{B}~\neq~x_{B}'$ can be written as
\begin{align}
\!\!\!\!\mathcal{P} \!\!\left\{ (x_{A}, x_{B}) \!\!\to\!\! (x_{A}, x_{B}')\right\}\!\! \lesssim\!\! \dfrac{1}{M^{2}}\!\!\!\!\sum_{(x_{A}, x_{B}) \in \mathcal{X}^{2}} \sum_{x_{B} \neq x_{B}' \in \mathcal{X}} \!\!\!\!\!\!\!\left( \Theta_{B, 1} + \Theta_{B, 2}\right),\!\!\!\! \label{error_case2}
\end{align}
where $\Theta_{B, 1} \triangleq \mathbb{E}[\Upsilon_{1}]/12$ and $\Theta_{B, 2} \triangleq \mathbb{E}[\Upsilon_{2}]/4$. The closed-form expression for $\Theta_{B, 1}$ can be obtained by replacing $N_A, E_A$ and $\Delta x_A$ by $N_B, E_B$ and $\Delta x_B$, respectively, in~\eqref{ThetaA1}. The closed-form expression for $\Theta_{B_2}$ can be obtained in a similar fashion using~\eqref{ThetaA2}.


\subsubsection*{Case III -- When $x_{A} \neq x_{A}'$, $x_{B} \neq x_{B}'$ and $\mathcal{M}_{c}(x_{A}, x_{B}) \neq \mathcal{M}_{c}(x_{A}', x_{B}')$}
This case is possible only for $M > 2$, because for the case of binary modulation (e.g., BPSK), if $x_{A} \neq x_{A}'$ and $x_{B} \neq x_{B}'$, both $(x_{A}, x_{B})$ and $(x_{A}^{\prime}, x_{B}^{\prime})$ will lie in the \emph{same} cluster for fixed network coding, i.e., $\mathcal{M}_{c}(x_{A}, x_{B}) = \mathcal{M}_{c}(x_{A}^{\prime}, x_{B}^{\prime})$ and hence a confusion among these pairs will not cause a symbol error event. Using \eqref{ChianiApp}, $\mathbb{E}[\Upsilon_{1}]$ for the given case can be written as
\begin{align}
	\mathbb{E}[\Upsilon_{1}]\! = & \mathbb{E} \left[ \exp \left( - \dfrac{1}{4N_{0}} \left\Vert \sqrt{E_{A}}\boldsymbol{h}_{A} \Delta x_{A} + \sqrt{E_{B}}\boldsymbol{h}_{B} \Delta x_{B}\right\Vert^{2}\right) \right] \notag \\
	= & \mathbb{E} \left[ \exp \left(-\dfrac{1}{4N_{0}} \sum_{j = 1}^{N_{R}} \left\vert \sqrt{E_{A}} h_{A, j} \Delta x_{A} + \sqrt{E_{B}} h_{B, j} \Delta x_{B}\right\vert^{2} \right) \right] \notag \\
	\leq & \mathbb{E} \left[ \exp \left\{ - \dfrac{1}{4N_{0}} \left( E_{A} |\Delta x_{A}|^{2} \sum_{j = 1}^{N_{R}} |h_{A, j}|^{2} + E_{B} |\Delta x_{B}|^{2} \sum_{j = 1}^{N_{R}}\!|h_{B, j}|^{2} \right.\right.\right. \notag \\
	& \hspace{6.5cm}\left.\left.\left.+ 2\sqrt{E_{A}E_{B}} \left\vert \Delta x_{A} \Delta x_{B}\right\vert \sum_{j = 1}^{N_{R}} \left\vert h_{A, j} h_{B, j} \right\vert \right) \right\} \right]. \label{Case3_difficult}
\end{align}
Since it is difficult in general to determine the PDF of $\sum_{j = 1}^{N_{R}}\left\vert h_{A, j}h_{B, j}\right\vert$, we consider here the analytically tractable case where $N_{R} = 1$. In this case, $\mathbb E (\Upsilon_1)$ is given by 
\begin{align}
	& \mathbb{E}(\Upsilon_{1}) = \mathbb{E} \left\{ \exp \left(-\dfrac{1}{4N_{0}} \left\vert \sqrt{E_{A}}h_{A} \Delta x_{A} + \sqrt{E_{B}}h_{B} \Delta x_{B}\right\vert^{2}\right)\right\} \nonumber \\
  = & \int_{0}^{\infty}\!\!\! \int_{0}^{\infty}\!\!\! \int_{-\pi}^{\pi}\!\!\exp\! \left(\!\dfrac{-E_{A}|\Delta x_{A}h_{A}|^{2}}{4N_{0}} \!\right)\! \exp \!\left(\!\dfrac{-E_{B} \left\vert \Delta x_{B}h_{B}\right\vert^{2}}{4N_{0}} \!\right)\notag \\
  & \hspace{3.5cm}\times \exp \left(\!\dfrac{-\sqrt{E_{A}E_{B}}\cos \theta |\Delta x_{A} \Delta x_{B}| |h_{A} h_{B}|}{2N_{0}} \!\right) \!f(\theta)f(|h_{A}|)f(|h_{B}|) d\theta d|h_{A}| d|h_{B}| \notag \\
  \myeqa & \int_{0}^{\infty} \exp \left(-\dfrac{E_{A}|\Delta x_{A}h_{A}|^{2}}{4N_{0}} \right) \int_{0}^{\infty} \exp \left(-\dfrac{E_{B}|\Delta x_{B}h_{B}|^{2}}{4N_{0}} \right) \notag \\
  & \hspace{5cm} \times \left[ I_{0}\left(\dfrac{\sqrt{E_{A}E_{B}}|\Delta x_{A} \Delta x_{B}|}{2N_{0}} |h_{A} h_{B}|\right)\right] f(|h_{A}|) f(|h_{B}|) d|h_{A}| d|h_{B}|\nonumber \\
  \myeqb & \int_{0}^{\infty} \exp \left(-\dfrac{E_{A}|\Delta x_{A}h_{A}|^{2}}{4N_{0}} \right) \sum_{l = 1}^{N_{B}} \binom{N_{B}}{l} (-1)^{l-1} \left[l \int_{0}^{\infty}2|h_{B}| \exp \left\{-\left(l + \dfrac{E_{B}|\Delta x_{B}|^{2}}{4N_{0}}\right)|h_{B}|^{2}\right\}  \right.  \notag \\
  & \hspace{6cm} \times \left. I_{0}\!\left(\dfrac{\sqrt{E_{A}E_{B}}|\Delta x_{A}\Delta x_{B}|}{2N_{0}} |h_{A}||h_{B}|\right) d|h_{B}| \right] f(|h_{A}|) d|h_{A}| \notag \\
  \myeqc & \int_{0}^{\infty} \!\!\exp \left(-\dfrac{E_{A}|\Delta x_{A}h_{A}|^{2}}{4N_{0}} \right) \sum_{l = 1}^{N_{B}} \binom{N_{B}}{l} (-1)^{l-1} (\Psi_{B,l})^{-1} \notag \\
  & \hspace{5cm} \times \exp \left[\dfrac{1}{4l \Psi_{B,l}} \left(\dfrac{\sqrt{E_{A}E_{B}}|\Delta x_{A}||\Delta x_{B}|}{2N_{0}} |h_{A}|\right)^{2}\right] f(|h_{A}|) d|h_{A}|  \nonumber \\
  = & \sum_{k=1}^{N_{A}} \sum_{l=1}^{N_{B}} \binom{N_{A}}{k} \binom{N_{B}}{l} (-1)^{(k+l-2)}  (\Psi_{B,l})^{-1} k \notag \\
  & \hspace{3.5cm} \times \int_{0}^{\infty} 2|h_{A}| \exp \!\left[\!-\!\left\{\underbrace{\!\!\dfrac{E_{A}|\Delta x_{A}|^{2}}{4N_{0}}\! -\! \dfrac{\left(\dfrac{\sqrt{E_{A}E_{B}}\left\vert\Delta x_{A}\Delta x_{B}\right\vert}{2N_{0}}\right)^{2}}{4l \Psi_{B,l}}}_{\eta}\! + k \!\right\}\! \left\vert h_{A} \right\vert^{2}\! \right] d \!\left\vert h_{A} \right\vert \nonumber \\
	= & \sum_{k=1}^{N_{A}} \sum_{l=1}^{N_{B}} \binom{N_{A}}{k} \binom{N_{B}}{l} \dfrac{(-1)^{(k+l-2)} (\Psi_{B,l})^{-1}}{1+\eta/k} \notag \\
	= &  \sum_{k=1}^{N_{A}} \sum_{l=1}^{N_{B}} \binom{N_{A}}{k} \binom{N_{B}}{l} (-1)^{(k+l-2)}\left(\Psi_{A,k}\Psi_{B,l} - \dfrac{1}{4kl}\Omega_{A,B}\right)^{-1}, \label{Case3_Upsilon1}
\end{align}
where $\theta = \angle h_{A} - \angle h_{B}$ is a random variable uniformly distributed over $[-\pi, \pi)$, (a) holds due to the fact that $\exp(\cos\theta)$ is an even function of $\theta$ and the integration w.r.t $\theta$ is solved using~\cite[p.~376]{Stegun}, where $I_{0}(\cdot)$ is the modified Bessel function of the first kind;~(b) is obtained from~(a) using the fact that the PDF $f(|h_{m}|), \, m \in \{A, B\}$ can be found by putting $N_{R} = 1$ in \eqref{fz}. Then the inner integral in~(b) is solved using~\cite[p.~306]{PrudnikovVol2}, yielding (c). Furthermore, $\Psi_{A, k}, \Psi_{B, l}$ and $\Omega_{A, B}$ in~\eqref{Case3_Upsilon1} are defined as $\Psi_{A, k} = 1 + \frac{E_{A} |\Delta x_{A}|^{2}}{4kN_{0}}$, $\Psi_{B, l} = 1 + \frac{E_{B} |\Delta x_{B}|^{2}}{4lN_{0}}$ and $\Omega_{A,B} = \left(\frac{\sqrt{E_{A}E_{B}}|\Delta x_{A} \Delta x_{B}|}{2N_{0}}\right)^{2}$.

Solving in the same fashion for $\mathbb{E}(\Upsilon_{2})$, we obtain
\begin{align}
	\mathbb{E}(\Upsilon_{2}) = & \sum_{k=1}^{N_{A}} \sum_{l=1}^{N_{B}} \binom{N_{A}}{k} \binom{N_{B}}{l} \dfrac{(-1)^{(k+l-2)}}{\left(\Xi_{A,k}\Xi_{B,l} - \dfrac{1}{4kl}\Phi_{A,B}\right)} ,\notag
\end{align}
\noindent where $\Xi_{A, k} = 1 + \frac{E_{A} |\Delta x_{A}|^{2}}{3kN_{0}}$, $\Xi_{B, l} = 1 + \frac{E_{B} |\Delta x_{B}|^{2}}{3lN_{0}}$ and $\Phi_{A,B} = \left(\frac{2\sqrt{E_{A}E_{B}}|\Delta x_{A} \Delta x_{B}|}{3N_{0}}\right)^{2}$.
Hence, the average SER arising from the case when $x_{A} \neq x_{A}^{\prime}$, $x_{B} \neq x_{B}^{\prime}$ and $\mathcal{M}_{c}(x_{A}, x_{B})~\neq~\mathcal{M}_{c}(x_{A}^{\prime}, x_{B}^{\prime})$ becomes equal to 
\begin{align}
	& P_{e}\left\{(x_{A}, x_{B}) \to (x_{A}^{\prime}, x_{B}^{\prime}), \mathcal{M}_{c}(x_{A}, x_{B}) \neq \mathcal{M}_{c}(x_{A}^{\prime}, x_{B}^{\prime})\right\} \nonumber \\
	\lesssim & \dfrac{1}{M^{2}} \sum_{(x_{A}, x_{B}) \in \mathcal{X}^{2}} \sum_{\substack{x_{A} \neq x_{A}^{\prime} \in \mathcal{X}, x_{B}\neq x_{B}^{\prime} \in \mathcal{X} \\ \mathcal{M}_{c}(x_{A}, x_{B}) \neq \mathcal{M}_{c}(x_{A}^{\prime}, x_{B}^{\prime})}} \sum_{k=1}^{N_{A}}  \sum_{l=1}^{N_{B}} \binom{N_{A}}{k} \binom{N_{B}}{l} (-1)^{(k+l-2)} \notag \\
	& \hspace{6cm}\times \left\{ \dfrac{1}{12}\left(\Psi_{A,k}\Psi_{B,l} - \dfrac{\Omega_{A,B}}{4kl}\right)^{-1} + \dfrac{1}{4}\left(\Xi_{A,k}\Xi_{B,l} - \dfrac{\Phi_{A,B}}{4kl}\right)^{-1}  \right\} \nonumber \\
	= & \dfrac{1}{M^{2}} \sum_{(x_{A}, x_{B}) \in \mathcal{X}^{2}} \sum_{\substack{x_{A} \neq x_{A}^{\prime} \in \mathcal{X}, x_{B}\neq x_{B}^{\prime} \in \mathcal{X} \\ \mathcal{M}_{c}(x_{A}, x_{B}) \neq \mathcal{M}_{c}(x_{A}^{\prime}, x_{B}^{\prime})}}  (\xi_{1} + \xi_{2}). \label{Case3_final}
\end{align}
where
\begin{equation}
	\xi_{1} = \sum_{k=1}^{N_{A}}  \sum_{l=1}^{N_{B}} \binom{N_{A}}{k} \binom{N_{B}}{l} \dfrac{(-1)^{(k+l-2)} }{12 \left(\Psi_{A,k}\Psi_{B,l} - \dfrac{\Omega_{A,B}}{4kl}\right)}, \label{Xi1}
\end{equation}
\begin{equation}	
	\xi_{2} = \sum_{k=1}^{N_{A}}  \sum_{l=1}^{N_{B}} \binom{N_{A}}{k} \binom{N_{B}}{l} \dfrac{(-1)^{(k+l-2)} }{4 \left(\Xi_{A,k}\Xi_{B,l} - \dfrac{\Phi_{A,B}}{4kl}\right)}. \notag
\end{equation}
An upper bound on the average SER for the MIMO-PNC system with AS1 for the special case of $N_{R} = 1$ can be obtained by putting $N_{R} = 1$ in \eqref{error_case1} and \eqref{error_case2}, and then adding \eqref{error_case1}, \eqref{error_case2} and \eqref{Case3_final}.
\subsubsection{Diversity Analysis}
To determine the diversity order of the MIMO-PNC system with AS1, we analyze the asymptotic decay rates of all three terms (i.e., corresponding to the three cases discussed above) on the right-hand side of \eqref{Pe_exact} separately. Considering first Case I ($x_A \neq x_A'$ and $x_B = x_B'$), using \eqref{pdf_z_tau} and \eqref{X2}, the PDF of $f(\mathfrak{z}_{A})$ can be rewritten as
\begin{align}
	f(\mathfrak{z}_{A}) \!=\! \dfrac{N_{A}\mathfrak{z}_{A}^{N_{R} - 1} \exp(-\mathfrak{z}_{A})}{(N_{R} - 1)!} \left[\! 1 \!-\! \exp (-\mathfrak{z}) \sum_{j = 0}^{N_{R} - 1} \dfrac{\mathfrak{z}^{j}}{j!}\right]^{N_{A} - 1}. \notag
\end{align}

Note that 
\begin{align}
	1 = & \exp(-\mathfrak{z}_{A}) \exp(\mathfrak{z}_{A}) = \sum_{j = 0}^{\infty} \dfrac{(-\mathfrak{z}_{A})^{j}}{j!} \sum_{l = 0}^{\infty} \dfrac{\mathfrak{z}_{A}^{l}}{l!} = \sum_{j = 0}^{\infty} \dfrac{(-\mathfrak{z}_{A})^{j}}{j!} \sum_{l = 0}^{N_{R} - 1} \dfrac{\mathfrak{z}_{A}^{l}}{l!} + \sum_{j = 0}^{\infty} \dfrac{(-\mathfrak{z}_{A})^{j}}{j!} \sum_{l = N_{R}}^{\infty} \dfrac{\mathfrak{z}_{A}^{l}}{l!}, \notag
\end{align}
and therefore,
\begin{align}	
	1 - \sum_{j = 0}^{\infty} \dfrac{(-\mathfrak{z}_{A})^{j}}{j!} \sum_{l = 0}^{N_{R} - 1} \dfrac{\mathfrak{z}_{A}^{l}}{l!} = \dfrac{\mathfrak{z}_{A}^{N_{R}}}{N_{R}!} + \mathcal{O}(\mathfrak{z}_{A}^{N_{R}}). \notag
\end{align}
Also we have $$\lim_{\mathfrak{z} \rightarrow 0^{+}}\exp(-\mathfrak{z}_{A}) = \lim_{\mathfrak{z}_{A} \rightarrow 0^{+}} \left(1 + \sum_{j = 1}^{\infty}\frac{(-\mathfrak{z}_{A})^{j}}{j!}\right) = \mathcal{O}(1).$$ Hence for $\mathfrak{z}_{A} \rightarrow 0^{+}$ we have
\begin{align}
	f(\mathfrak{z}_{A}) & = \dfrac{N_{A}}{(N_{R} - 1)!} \left[ \dfrac{\mathfrak{z}_{A}^{N_{R}}}{N_{R}!} + \mathcal{O}\left(\mathfrak{z}_{A}^{N_{R}}\right)\right]^{N_{A} - 1} \mathfrak{z}_{A}^{N_{R} - 1} \mathcal{O}(1) = \dfrac{N_{A}}{(N_{R} - 1)!} \dfrac{\mathfrak{z}_{A}^{N_{A}N_{R} - 1}}{(N_{R}!)^{N_{A} - 1}} + \mathcal{O}\left(\mathfrak{z}_{A}^{N_{A}N_{R} - 1}\right). \notag
\end{align}
Note that for the case when $x_A \neq x_A'$ and $x_B = x_B'$ (i.e.,~Case~I), user $B$ can be assumed to be absent (as the transmission from user $B$ do not cause any error at the relay) and the PNC system reduces to simple single transmitter -- single receiver setting. Therefore, using~\cite[\textit{Proposition 1}]{Giannakis}, we can conclude that $\Theta_{A, 1}$ and $\Theta_{A, 2}$ decay as $(E_{A}/N_{0})^{-N_{A} N_{R}}$ for large values of $E_{A}/N_{0}$, and hence the diversity order of the term arising from Case I is $N_AN_R$. Following the same argument, $\Theta_{B, 1}$ and $\Theta_{B, 2}$ decay as $(E_{B}/N_{0})^{-N_{B} N_{R}}$ for large values of $E_{B}/N_{0}$, the diversity order of the term arising from Case II is $N_B N_R$.

For Case III with $N_R = 1$, substituting the values of $\Psi_{A,k}$, $\Psi_{B,l}$ and $\Omega_{A,B}$ into \eqref{Xi1}, $\xi_{1}$ becomes (where we define $E_{\min} \triangleq \min(E_{A}, E_{B})$)
\begin{align}
	\xi_{1} = & \sum_{k=1}^{N_{A}}  \sum_{l=1}^{N_{B}} \binom{N_{A}}{k} \binom{N_{B}}{l} \times \dfrac{(-1)^{(k+l-2)}} {12 \left(1 + \dfrac{E_{A}|\Delta x_{A}|^{2}}{4kN_{0}} + \dfrac{E_{B}|\Delta x_{B}|^{2}}{4lN_{0}}\right)} \nonumber \\
	\leq & \sum_{k=1}^{N_{A}}  \sum_{l=1}^{N_{B}} \binom{N_{A}}{k} \binom{N_{B}}{l}  \dfrac{(-1)^{(k+l-2)}} {12 \left(1 + \dfrac{E_{\min}|\Delta x_{A}|^{2}}{4kN_{0}} + \dfrac{E_{\min}|\Delta x_{B}|^{2}}{4lN_{0}}\right)} \notag  \\
	= & \dfrac{1}{12}\sum_{k=1}^{N_{A}}  \sum_{l=1}^{N_{B}} \binom{N_{A}}{k} \binom{N_{B}}{l}\dfrac{(-1)^{(k+l-2)}  \left(g(k, l)\frac{E_{\min}}{N_{0}}\right)^{-1}}{\left(1 + \dfrac{1}{g(k, l) \frac{E_{\min}}{N_{0}}}\right)}, \notag
\end{align}
where $g(k, l) = \frac{|\Delta x_{A}|^{2}}{4k} + \frac{|\Delta x_{B}|^{2}}{4l}$.
Using the binomial expansion, $\xi_{1}$ can be rewritten as
\begin{align}
	\xi_{1} \leq & \dfrac{1}{12}\sum_{k=1}^{N_{A}}  \sum_{l=1}^{N_{B}} \binom{N_{A}}{k} \binom{N_{B}}{l} \dfrac{(-1)^{(k+l-2)}}{g(k, l) \frac{E_{\min}}{N_{0}}}  \sum_{m = 1}^{\infty} \left(\dfrac{-1}{g(k, l)\frac{E_{\min}}{N_{0})}}\right)^{m - 1} \nonumber  \\	
	= & \sum_{m = 1}^{\infty} \left(\dfrac{E_{\min}}{N_{0}}\right)^{-m} \underbrace{\dfrac{1}{12}\sum_{k=1}^{N_{A}}  \sum_{l=1}^{N_{B}} \binom{N_{A}}{k} \binom{N_{B}}{l}  (-1)^{k+l+m-3}(g(k, l))^{-m}}_{B_{-m}} \nonumber \\
	= &\sum_{m = 1}^{\infty} B_{-m} \left(\dfrac{E_{\min}}{N_{0}}\right)^{-m} \!<\! B_{-1} \left(\dfrac{E_{\min}}{N_{0}}\right)^{-1} + \mathcal{O} \left[\left(\dfrac{E_{\min}}{N_{0}}\right)^{-1}\right]. \label{xi1_order}
\end{align}

Similarly, for $\xi_{2}$ it can be shown that
\begin{align}
	\xi_{2} < B_{-1}^{\prime} \left(\dfrac{E_{\min}}{N_{0}}\right)^{-1} + \mathcal{O} \left[\left( \dfrac{E_{\min}}{N_{0}}\right)^{-1} \right] \label{xi2_order}
\end{align}
where
\begin{equation}
	B_{-m}^{\prime} = \dfrac{1}{4}\sum_{k=1}^{N_{A}}  \sum_{l=1}^{N_{B}} \binom{N_{A}}{k} \binom{N_{B}}{l}  (-1)^{k+l+m-3}(g^{\prime}(k, l))^{-m}, \nonumber
\end{equation}
\begin{equation}
	g^{\prime}(k, l) = \dfrac{|\Delta x_{A}|^{2}}{3k} + \dfrac{|\Delta x_{B}|^{2}}{3l}. \nonumber
\end{equation}

From \eqref{Case3_final}, \eqref{xi1_order} and \eqref{xi2_order}, it is clear that the average symbol error probability due to the case when $N_{R} = 1$, $x_{A} \neq x_{A}^{\prime}$, $x_{B} \neq x_{B}^{\prime}$ and $\mathcal{M}_{c}\left(x_{A}, x_{B}\right) \neq \mathcal{M}_{c} \left(x_{A}^{\prime}, x_{B}^{\prime}\right)$ decays as $(E_{\min}/N_{0})^{- 1}$ for higher values of $E_{\min}/N_{0}$. 

The discussion above regarding the asymptotic decay rates of $\Theta_{A, 1}, \Theta_{A, 2}, \Theta_{B, 1}, \Theta_{B, 2}, \xi_1$ and $\xi_2$ leads to a couple of important observations, which are listed below:
\begin{enumerate}[a)]
	\item The first term in the right-hand side of \eqref{Pe_exact} decays as $(E_{A}/N_{0})^{-N_{A}N_{R}}$ for higher values of $E_{A}/N_{0}$ while the second term decays as $(E_{B}/N_{0})^{-N_{B}N_{R}}$ for higher values of $E_{B}/N_{0}$. It is important to recall that for binary modulations ($M = 2$) (as in~\cite{Huang}), the third term in the right-hand side of~\eqref{Pe_exact} (which is analyzed in \emph{Case III}) do not contribute to the symbol error probability at the relay and hence the overall system diversity order becomes $\min(N_{A}, N_{B}) \times N_{R}$. This completes the proof for the diversity order of the MIMO-PNC system with BPSK modulation.
	\item For the MISO-PNC system $(N_{R} = 1)$ with AS1 and non-binary modulations ($M > 2$), since the first two terms on the right-hand side of \eqref{Pe_exact} will decay as $(E_{A}/N_{0})^{-N_{A}}$ and $(E_{A}/N_{0})^{-N_{B}}$, respectively, for higher values of $E_{A}/N_{0}$ and $E_B/N_0$, respectively, while the third term will decay as $(E_{\min}/N_{0})^{- 1}$ for higher values of $E_{\min}/N_0$, the system diversity order becomes $\min(N_{A}, N_{B}, 1) = 1$, irrespective of the number of antennas at the users.
\end{enumerate}
In case of the MIMO-PNC system with AS1 and non-binary modulations, since it is difficult in general (as concluded from \eqref{Case3_difficult}) to represent/bound the average SER for \emph{Case III} in a tractable form, we present extensive simulation results (detailed in Section VI), which confirms that the system fails to achieve transmit diversity and the system diversity order drops to $N_{R}$.

The following section deals with the derivation of upper and lower bounds on the average SER of the MIMO-PNC system with AS2.

\section{AS2: Antenna selection based on ED metric}
To analyze the performance of AS2, we follow the approach adopted in~\cite{Hari} and~\cite{UpBound}. In~\cite{Hari} the authors analyzed the error performance of an ED based antenna selection scheme for SM. The analysis for PNC differs due to that fact that for an SM system, only a single antenna is active during transmission, while for a TWRC with PNC, two antennas (one from each user) transmit simultaneously. In~\cite{UpBound} the authors used a similar technique to analyze the error performance of an SMx system with ED based transmit antenna selection. 

Given a set of user-antenna indices $I = (i, j) \in \mathcal{I}$, the set of possible transmit vectors for the PNC system can be defined as $\mathcal{C}_{I} = \left\{\left[\sqrt{E_{A}}x_{A} \boldsymbol{e}_{i} \, \sqrt{E_{B}}x_{B} \boldsymbol{e}_{j}\right]^{T} | x_{A}, x_{B} \in \mathcal{S}\right\}$, where $\boldsymbol{e}_{i}$ and $\boldsymbol{e}_{j}$ are row vectors of length $N_{A}$ and $N_{B}$ respectively with all zero elements apart from a 1 at the $i^{\text{th}}$ and $j^{\text{th}}$ position, respectively. Let $\boldsymbol{z}_{I}(x_{A}, x_{B}) = \left[\sqrt{E_{A}}x_{A} \boldsymbol{e}_{i} \, \sqrt{E_{B}}x_{B}\boldsymbol{e}_{j}\right]^{T}$.

\subsubsection{Lower bound on diversity order}
Defining $\Delta \mathcal{C}_{I} = \Big\{\boldsymbol{z}_{I} \left(x_{A}^{(1)}, x_{B}^{(1)}\right) - \boldsymbol{z}_{I} \left(x_{A}^{(2)}, x_{B}^{(2)}\right) \Big| x_{A}^{(1)}, x_{B}^{(1)}, x_{A}^{(2)}$, $x_{B}^{(2)} \in \mathcal{S}$, $\mathcal{M}_{c}\left(x_{A}^{(1)}, x_{B}^{(1)}\right) \neq \mathcal{M}_{c}\left(x_{A}^{(2)}, x_{B}^{(2)}\right) \Big\}$ as the set of difference vectors corresponding to the codebook $\mathcal{C}_{I}$, the set of matrices $\Delta \mathcal{D}$ can be defined as
\begin{equation}
	\Delta \mathcal{D} = \{[\boldsymbol{x}_{1}, \boldsymbol{x}_{2}, \ldots, \boldsymbol{x}_{n}]|\boldsymbol{x}_{k} \in \Delta \mathcal{C}_{k} \forall k \in \{1, 2, \ldots, n\}\}. \notag
\end{equation}
Each element in $\Delta \mathcal{D}$ will be of size $(N_{A} + N_{B}) \times n$ where $n = N_{A} \times N_{B}$. Let $r_{\text{min}}$ be defined as 
\begin{equation}
	r_{\text{min}} \triangleq \min \{ \mathrm{rank}(\boldsymbol{X})| \boldsymbol{X} \in \Delta \mathcal{D}\}. \notag
\end{equation}
The minimum number of \emph{linearly independent} columns in $\boldsymbol{X}$, i.e. $r_{\text{min}}$ will be $\min\{N_{A}, N_{B}\}$. To understand this, consider an example when $N_{A} = 3 $ and $N_{B} = 2$. The structure of each element in $\Delta \mathcal{D}$ is given by 
\begin{equation}
	\boldsymbol{X} = \begin{bmatrix}
	\sqrt{E_{A}}\Delta x_{A}^{(1)} & \sqrt{E_{A}}\Delta x_{A}^{(2)} & 0 & 0 & 0 & 0 \\
	0 & 0 & \sqrt{E_{A}}\Delta x_{A}^{(3)} & \sqrt{E_{A}}\Delta x_{A}^{(4)} & 0 & 0 \\
	0 & 0 & 0 & 0 & \sqrt{E_{A}}\Delta x_{A}^{(5)} & \sqrt{E_{A}}\Delta x_{A}^{(6)} \\ 
	\sqrt{E_{B}}\Delta x_{B}^{(1)} & 0 & \sqrt{E_{B}}\Delta x_{B}^{(3)} & 0 & \sqrt{E_{B}}\Delta x_{B}^{(5)} & 0\\
	0 & \sqrt{E_{B}}\Delta x_{B}^{(2)} & 0 & \sqrt{E_{B}}\Delta x_{B}^{(4)} & 0 & \sqrt{E_{B}}\Delta x_{B}^{(6)}
	\end{bmatrix}. \label{X_structure}
\end{equation}

Thanks to definition of $\Delta \mathcal{C}_{I}$, in which the condition $\mathcal{M}_{c}\left(x_{A}^{(1)}, x_{B}^{(1)}\right) \neq \mathcal{M}_{c}\left(x_{A}^{(2)}, x_{B}^{(2)}\right)$ ensures that $\boldsymbol{z}_{I} \left(x_{A}^{(1)}, x_{B}^{(1)}\right) \neq \boldsymbol{z}_{I} \left(x_{A}^{(2)}, x_{B}^{(2)}\right)$, implies that in any of the columns of $\boldsymbol{X}$, both $\Delta x_{A}^{(i)}$ and $\Delta x_{B}^{(i)}$ cannot be zero simultaneously. If $\Delta x_{B}^{(1)}$ and $\Delta x_{B}^{(2)}$ are non-zero, they form a non-zero minor (a diagonal matrix) and the minimum possible rank of $\boldsymbol{X}$ becomes 2. Now if $\Delta x_{B}^{(1)}$ is zero and any one from $\Delta x_{B}^{(3)}$ or $\Delta x_{B}^{(5)}$ is non-zero then also a $2 \times 2$ non-zero minor can be formed using $\Delta x_{B}^{(2)}$. A similar argument applies when $\Delta x_{B}^{(1)}$ is non-zero and $\Delta x_{B}^{(2)}$ is zero and a $2 \times 2$ minor can be formed using $\Delta x_{B}^{(1)}$ and $\Delta x_{B}^{(4)}$ or $\Delta x_{B}^{(6)}$ with the help of a column swap. In the case where $\Delta x_{B}^{(1)}$, $\Delta x_{B}^{(3)}$ and $\Delta x_{B}^{(5)}$ are zero, the $\Delta x_{A}^{(i)}$ values in the corresponding columns will be non-zero and they will form three linearly independent columns and the rank of matrix $\boldsymbol{X}$ will be at least 3. A similar argument is valid for the case when $\Delta x_{B}^{(2)}$, $\Delta x_{B}^{(4)}$ and $\Delta x_{B}^{(6)}$ are zero. On the other hand, if all the $\Delta x_{B}^{(i)}$ values are zero then the first three rows will be linearly independent and the rank of $\boldsymbol{X}$ will be 3. Hence the minimum possible rank of $\boldsymbol{X}$ is 2 i.e., $\min \{N_{A}, N_{B} \}$. It is straightforward to generalize this argument to the case of an arbitrary number of antennas at each user.

Let the transmit vectors in the each codebook be denoted as $\mathcal{C}_{k} = \left\{\boldsymbol{x}_{l}(k)|l \in \left\{1, 2, ,\ldots, M^{2}\right\}\right\}$ and the optimal set of user-antennas for any particular channel realization $\boldsymbol{H}$ be $I_{k^{\ast}}$. For $E_{\min}/N_{0} \gg 1$, the average pairwise error probability between any two different transmit vectors indexed by $l_{1}$ and $l_{2}$ in the codebook $\mathcal{C}_{k^{\ast}}$ can be expressed, using the Chernoff bound, as \cite[eq.~(4)-(10)]{Hari}
\begin{align}
	\mathbb{E}\left\{\mathcal{P}(\boldsymbol{x}_{l_{1}} \rightarrow \boldsymbol{x}_{l_{2}}) \right\} \leq & \dfrac{1}{2} \left(\dfrac{E_{\min}\lambda^{\ast}}{4nN_{0}} \right)^{-N_{R}r_{\text{min}}}, \notag
\end{align}
where $\lambda^{\ast} = \min_{\boldsymbol{X} \in \Delta \mathcal{D}} \lambda_{s}(\boldsymbol{XX}^{H})$ and $\lambda_{s}(\boldsymbol{Y})$ denotes the smallest non-zero eigenvalue of matrix $\boldsymbol{Y}$. An upper bound on the average SER for AS2 at $E_{\min}/N_{0} \gg 1$ can therefore be given as
\begin{align}
	P_{e} \leq & \ \dfrac{1}{2M^{2}} \sum_{\boldsymbol{x}_{l_{1}} \in \mathcal{C}_{k^\ast}} \!\!\!\sum_{\substack{\boldsymbol{x}_{l_{1}} \neq \boldsymbol{x}_{l_{2}} \in \mathcal{C}_{k^{\ast}} \\ \mathcal{M}_{c} \left(x_{A}^{(l_{1})}, x_{B}^{(l_{1})}\right) \neq \mathcal{M}_{c} \left(x_{A}^{(l_{2})}, x_{B}^{(l_{2})}\right)}}  \!\!\!\!\left(\dfrac{E_{\min}\lambda^{\ast}}{4nN_{0}} \right)^{-N_{R}r_{\text{min}}} = \binom{M}{2} \left(\dfrac{E_{\min}\lambda^{\ast}}{4nN_{0}} \right)^{-\min (N_{A}, N_{B}) \times N_{R}}. \label{Pe_ED}
\end{align}
It is clear from \eqref{Pe_ED} that the PNC system with AS2 achieves a diversity order lower bounded by $\min(N_{A},N_{B}) \times N_{R}$ for any modulation order $M$.

\subsubsection{Upper bound on diversity order}
The average pairwise error probability i.e., the probability that the signal pair $\boldsymbol{\tilde{x}}~=~(\sqrt{E_{A}}\tilde{x}_{A}, \sqrt{E_{B}}\tilde{x}_{B})$ is more likely than $\boldsymbol{x} = (\sqrt{E_{A}}x_{A}, \sqrt{E_{B}}x_{B})$ from the receiver's perspective can be given as
\begin{equation}
	\mathbb{E}\left\{ \mathcal{P}(\boldsymbol{x} \!\to \! \boldsymbol{\tilde{x}}) \right\}\! =\! \mathbb{E} \!\left\{ \!\dfrac{1}{\pi}\! \int_{0}^{\pi/2} \!\!\!\!\exp \left( - \dfrac{\left\Vert \boldsymbol{H}_{I_{ED}} (\boldsymbol{x} - \boldsymbol{\tilde{x}})\right\Vert^{2}}{4N_{0} \sin^{2} \theta}\right)d\theta\right\}. \notag
\end{equation}

As defined earlier, the possible channel matrices are
\begin{equation}
	\begin{aligned}
		\boldsymbol{H}_{1} & = \left[ \boldsymbol{\mathfrak{h}}_{A,1}, \, \boldsymbol{\mathfrak{h}}_{B, 1}\right] \\
		\boldsymbol{H}_{2} & = \left[ \boldsymbol{\mathfrak{h}}_{A,2}, \, \boldsymbol{\mathfrak{h}}_{B, 1}\right] \\
		& \, \, \, \vdots \\
		\boldsymbol{H}_{n} & = \left[ \boldsymbol{\mathfrak{h}}_{A, N_{A}}, \, \boldsymbol{\mathfrak{h}}_{B, N_{B}}\right] \\
	\end{aligned} \notag
\end{equation}

In general, $\boldsymbol{H}_{i} = [\boldsymbol{\mathfrak{h}}_{A, \zeta}, \boldsymbol{\mathfrak{h}}_{B, \psi}]$, with $1 \leq i \leq n$, $1 \leq \zeta \leq N_{A}$, $1 \leq \psi \leq N_{B}$, $\boldsymbol{\mathfrak{h}}_{A, \xi} = [\mathfrak{h}_{A, \zeta, 1} \ \cdots \ \mathfrak{h}_{A, \zeta, N_{R}}]^{T}$ and $\boldsymbol{\mathfrak{h}}_{B, \psi} = [\mathfrak{h}_{B, \psi, 1} \ \cdots \ \mathfrak{h}_{B, \psi, N_{R}}]^{T}$. The diversity order with optimal user-antenna combination is written as
\begin{equation}
	d = \min_{\boldsymbol{x}, \boldsymbol{\tilde{x}} \neq \boldsymbol{x}} \left[ \lim_{N_{0} \rightarrow 0} \dfrac{\log \mathcal{P}(\boldsymbol{x} \rightarrow \boldsymbol{\tilde{x}})}{\log N_{0}}\right]. \label{DivOrder}
\end{equation}
For a signal pair $(\boldsymbol{x}, \boldsymbol{\tilde{x}})$ that satisfies $x_{A} \neq \tilde{x}_{A}$ and $x_{B} = \tilde{x}_{B}$, it follows from~\eqref{DivOrder} that 
\begin{equation}
	d \leq \lim_{N_{0} \rightarrow 0} \dfrac{\log \mathcal{P}(\boldsymbol{x} \rightarrow \boldsymbol{\tilde{x}})}{\log N_{0}}. \notag
\end{equation}
Furthermore, 
\begin{align}
	\left\Vert \boldsymbol{H}_{I_{ED}} (\boldsymbol{x} - \boldsymbol{\tilde{x}})\right\Vert^{2} & \leq \max_{1 \leq i \leq n} \left\Vert \boldsymbol{H}_{i} (\boldsymbol{x} - \boldsymbol{\tilde{x}})\right\Vert^{2} = \sqrt{E_A}\,|x_{A} - \tilde{x}_{A}|^{2} \max_{1 \leq \zeta \leq N_{A}} \left\Vert \boldsymbol{\mathfrak{h}}_{A, \zeta}\right\Vert^{2} \notag \\
	& \leq \sqrt{E_{A}}\,|x_{A} - \tilde{x}_{A}|^{2} \sum_{\zeta = 1}^{N_{A}} \sum_{k = 1}^{N_{R}} |h_{A, \zeta, k}|^{2}. \notag
\end{align}
Using the arguments in \cite[Section III-i]{UpBound}, it follows that 
\begin{equation}
	d \leq N_{A} \times N_{R}. \label{dA_UB_value}
\end{equation}
Similarly, for the signal pair $(\boldsymbol{x}, \boldsymbol{\tilde{x}})$ that satisfies $x_{A} = \tilde{x}_{A}$ and $x_{B} \neq \tilde{x}_{B}$, the diversity order can be upper bound by 
\begin{equation}
	d \leq N_{B} \times N_{R}. \label{dB_UB_value}
\end{equation}
Using \eqref{dA_UB_value} and \eqref{dB_UB_value}, the upper bound on the diversity order for MIMO-PNC with AS2 can be given by $\min(N_{A}, N_{B}) \times N_{R}$ for any modulation order $M$. 

Since both upper and lower bounds on the diversity order for MIMO-PNC with AS2 are equal, we may conclude that the exact diversity order for the MIMO-PNC system is equal to $\min(N_{A}, N_{B}) \times N_{R}$ for any modulation order $M$.

\section{Results and Discussion}
In this section, we present a performance comparison of the two AS schemes discussed in the previous sections. 
\subsection{Simulation setup} 
For all Monte Carlo simulations, our setup is as follows. We generate random QPSK symbols $x_A, x_B \in \mathcal X$, and then compute $x_R = \mathcal{M}_c (x_A , x_B)$ using the PNC mapping shown in~Table~\ref{mapping_table}. Next, we generate i.i.d. random samples of $\mathfrak h_{m, i, j} \sim \mathcal{CN}(0, 1)$ for every $m \in \{A, B\}, 1 \leq i \leq N_m, 1 \leq j \leq N_R$. For AS1, the index of the optimal user-antenna is given by $i_m^* = \argmax_{1 \leq i \leq N_m} \sum_{j = 1}^{N_R} |\mathfrak h_{m, i, j}|^2$, whereas for AS2, the indices of the optimal user-antennas are obtained using~\eqref{index_AS2}. The noise vector $\bs n \in \mathbb C^{N_R \times 1}$ is then generated whose elements are independent and complex Gaussian with zero mean and variance $N_0$, and given $E_A$ and $E_B$ we obtain the signal vector received at the relay as shown in~\eqref{signal_received}. Finally, we compute the relay's ML estimate of the transmitted symbol pair $(\hat x_A, \hat x_B)$ using~\eqref{MLE}, and this is used to obtain the estimated network-coded symbol, denoted by $\hat x_R = \mathcal M_c ( \hat x_A, \hat x_B)$. The average SER is measured by counting the number of error events, i.e., $\hat x_R \neq x_R$, and dividing by the number of symbols transmitted.
\begin{figure}[t]
\centering
\begin{minipage}{.45\textwidth}
  \centering
  \includegraphics[scale=0.5]{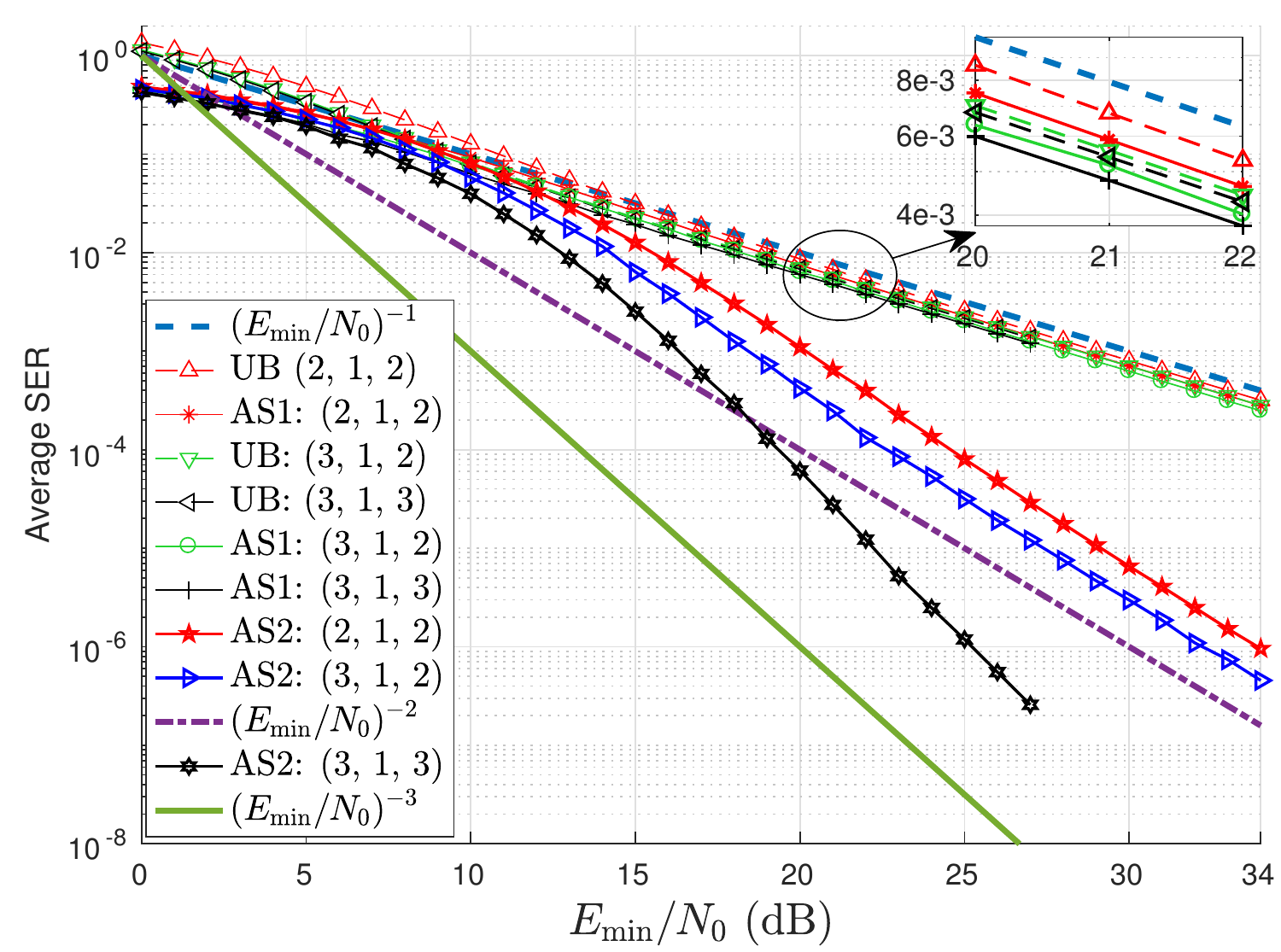}
  \caption{Average SER performance comparison for the two AS schemes in MISO case $(N_{R} = 1)$ in symmetric channels $(E_{A}/N_{0} = E_{B} / N_{0} = E_{\min} / N_{0})$.} 
  \label{MISO_Symm}
\end{minipage}%
\hfill
\begin{minipage}{.45\textwidth}
  \vspace{0.65cm}
  \centering
  \includegraphics[scale=0.5]{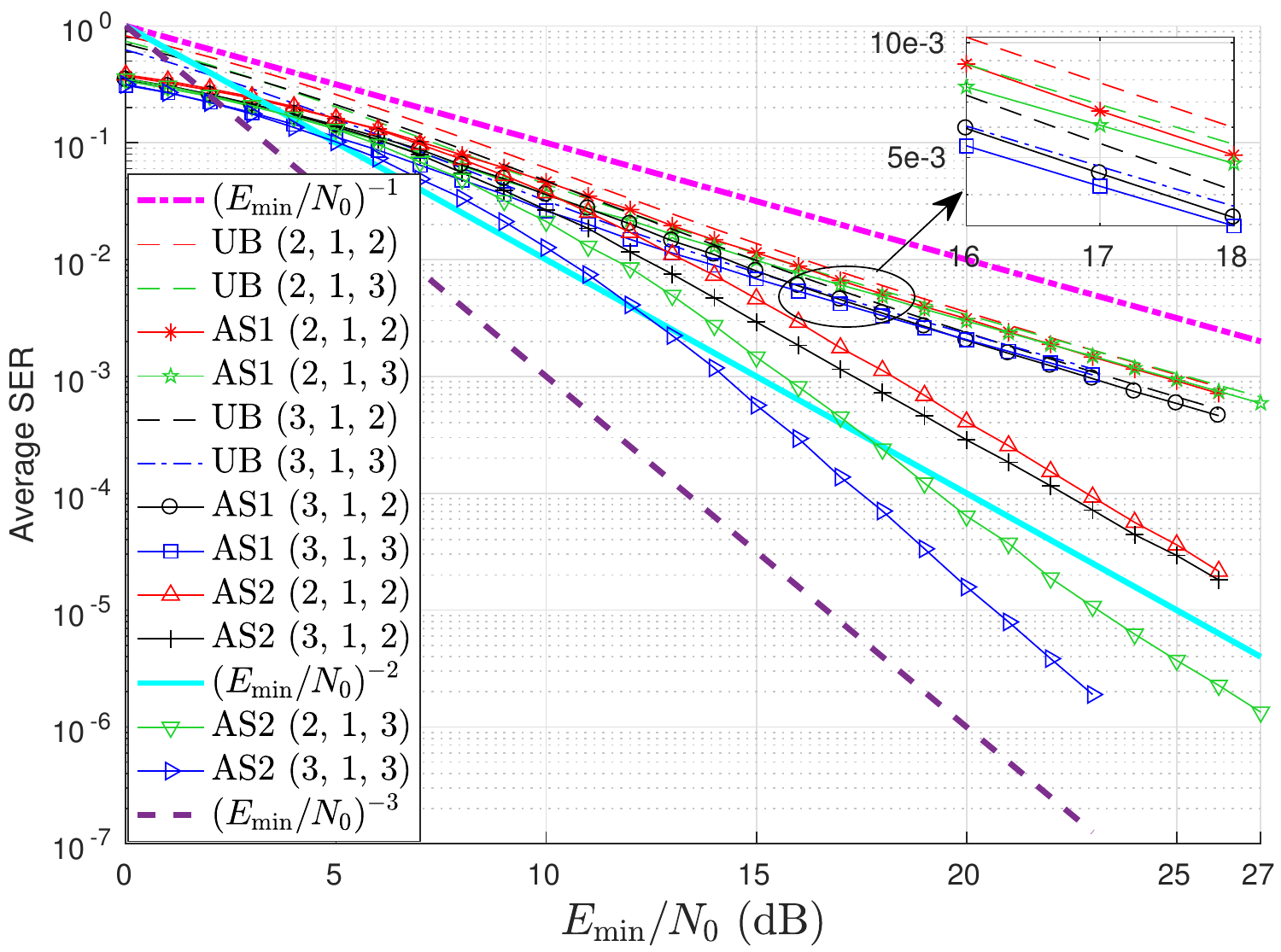}
  \caption{Average SER performance comparison for the two AS schemes in MISO case $(N_{R} = 1)$ in asymmetric channels, where $E_{A} / N_{0} = (E_{\min} / N_{0}) + 5 \text{ dB}$ and $E_{B} / N_{0} = E_{\min} / N_{0}$.} 
  \label{MISO_Assym}
\end{minipage}
\end{figure}

\subsection{Discussion} In Fig. \ref{MISO_Symm}, the SER performance for the two AS schemes is shown for the case when $N_{R} = 1$ with different number of antennas at the users in symmetric channels (i.e., $E_{A}/N_{0} = E_{B}/N_{0} = E_{\min} / N_{0}$). In the figure legend, `UB' denotes the upper bound on the average SER for AS1 (as derived in Section IV) and the numbers in parentheses denote $(N_{A}, N_{R}, N_{B})$. The plots marked `UB' have been drawn by substituting $N_{R} = 1$ in \eqref{error_case1} and \eqref{error_case2}, and then adding \eqref{error_case1}, \eqref{error_case2} and \eqref{Case3_final}, whereas the plots for AS1 and AS2 have been drawn using the Monte Carlo simulations (as described in the previous subsection). It is clear from the figure that for AS1 (where the user-antenna is selected based on the maximization of the overall channel gain between the user and the relay), the system diversity order becomes equal to 1 irrespective of the number of antennas at the users' end, as for higher values of $E_{\min}/N_{0}$ the SER curve becomes parallel to $(E_{\min}/N_{0})^{-1}$ in each case as was proved in Section IV. It is also worth noting that the derived closed-form expression for the upper bound on the SER is very tight for AS1. 

In contrast to this, the PNC system with AS2 (where the user antenna is selected based on the ED metric) outperforms the one with AS1 while achieving a higher diversity order. For the case when $N_{A} = N_{B} = 2$ and $N_{A} = 3$, $N_{B} = 2$, the average SER in the PNC system with AS2 decays more rapidly as compared to AS1 and becomes parallel to $(E_{\min}/N_{0})^{-\min(N_{A}, N_{B}) \times N_{R}} = (E_{\min}/N_{0})^{-2}$ for higher values of $E_{\min}/N_{0}$ as was proved in Section V. Similarly, for the case when $N_{A} = N_{B} = 3$ the average SER for AS1 decays as $(E_{\min}/N_{0})^{-1}$ while for AS2 the average SER decays as $(E_{\min}/N_{0})^{-3}$ at higher values of $E_{\min}/N_{0}$.

Fig. \ref{MISO_Assym} shows the average SER performance comparison of the two AS schemes in asymmetric channels with different number of user antennas and $N_{R} = 1$. In this case $E_{A} / N_{0} = (E_{\min} / N_{0}) + 5$ dB (and thus $E_{B} / N_{0} = E_{\min} / N_{0}$). Similar to the previous results, the diversity order achieved by the MISO-PNC system for asymmetric channels with AS1 is 1, while the system with AS2 achieves the full diversity order of $\min(N_{A}, N_{B})$ (recall that here $N_{R} = 1$).
\begin{figure}[t]
\centering
\begin{minipage}{.45\textwidth}
  \centering
  \includegraphics[scale=0.5]{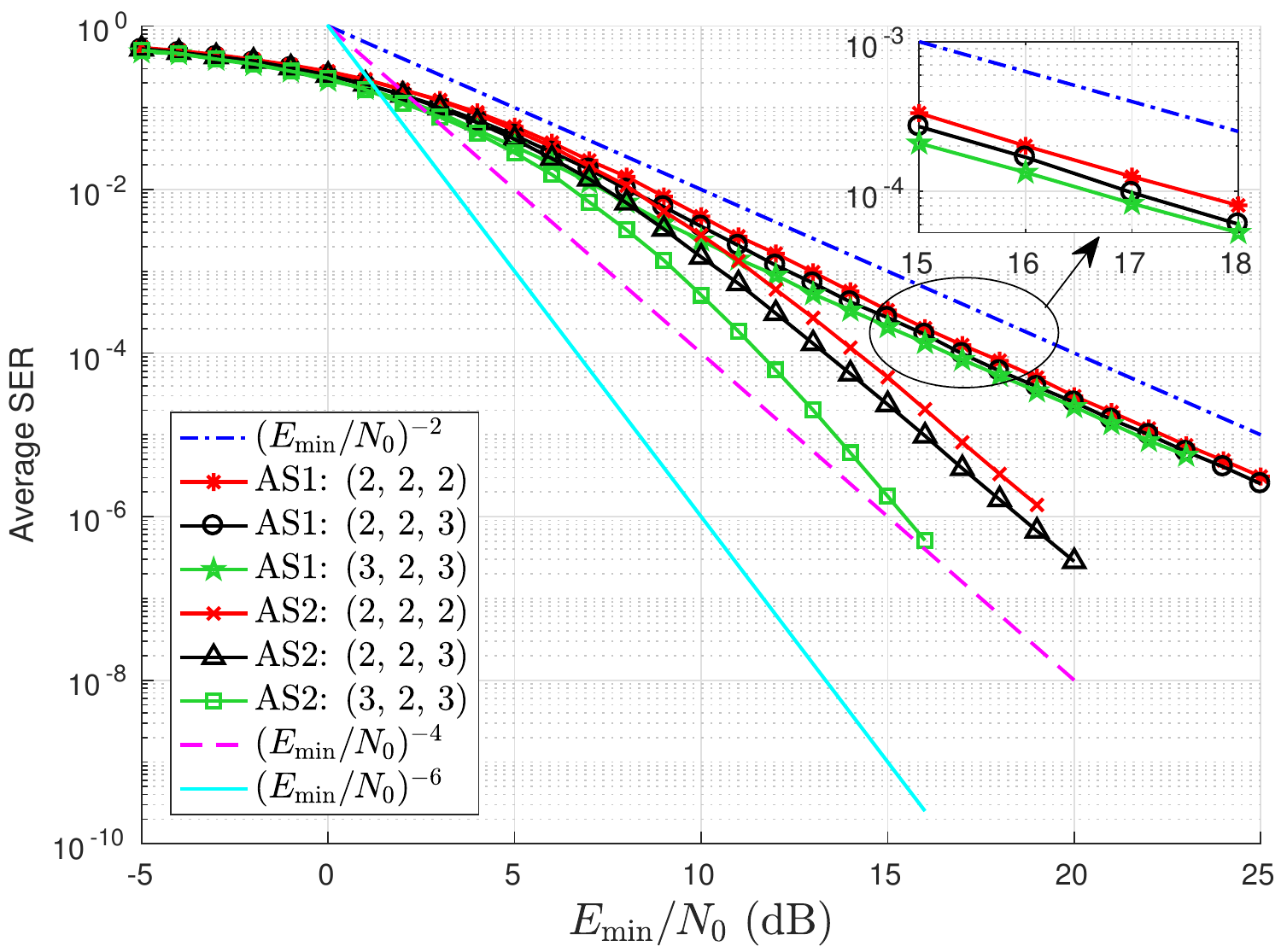}
  \caption{Average SER performance comparison for the two AS schemes in MIMO case $(N_{R} > 1)$ with symmetric channels, where $E_{A}/N_{0} = E_{B} / N_{0} = E_{\min} / N_{0}$).} 
  \label{MIMO_Symm}
\end{minipage}%
\hfill
\begin{minipage}{.45\textwidth}
  \vspace{0.65cm}
  \centering
  \includegraphics[scale=0.5]{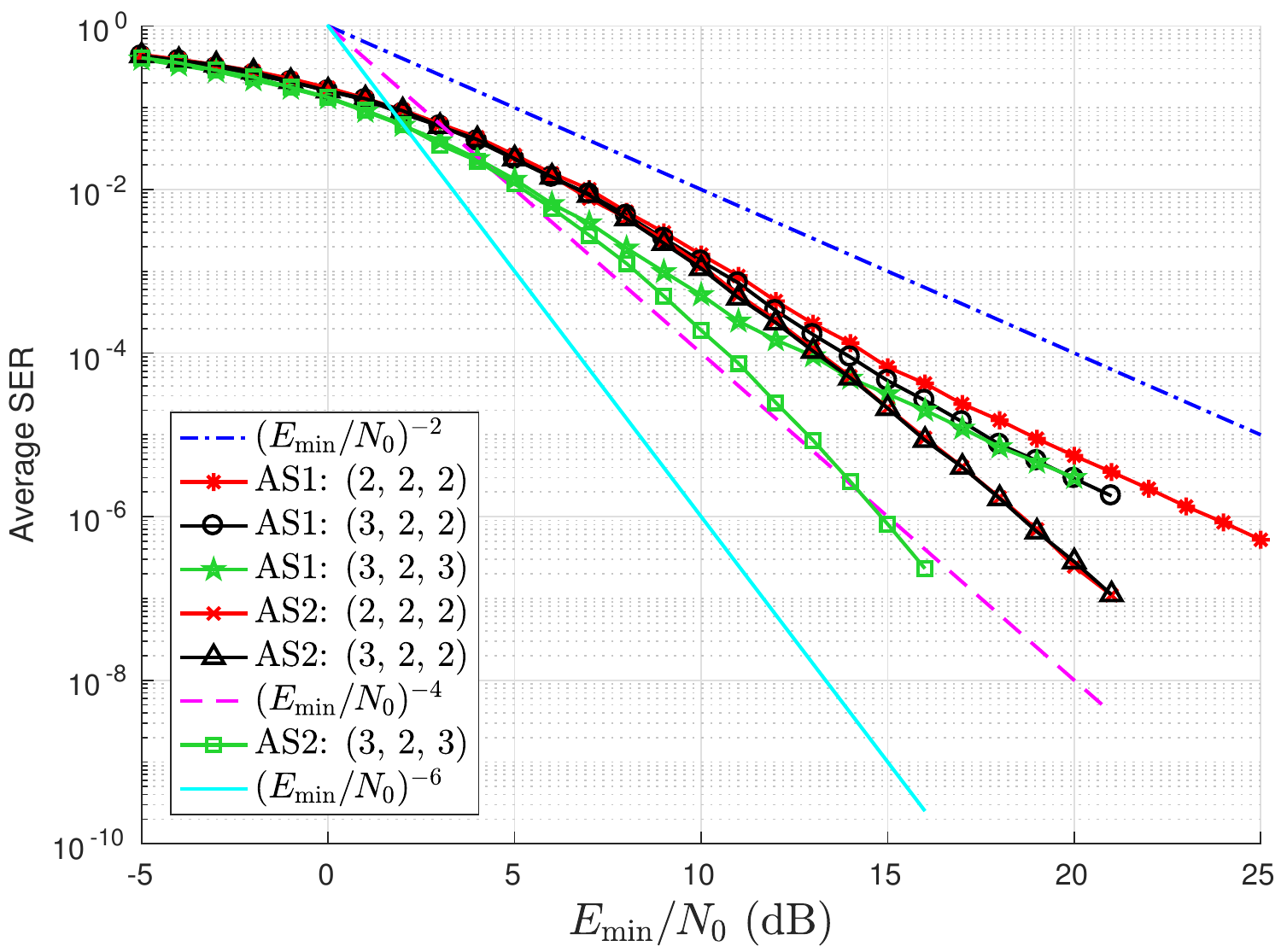}
  \caption{Average SER performance comparison for the two AS schemes in MIMO case $(N_{R} > 1)$ with asymmetric channels, where $E_{A}/N_{0} = (E_{\min}/N_{0}) + 5$ dB and $E_{B} / N_{0} = E_{\min} / N_{0}$).} 
  \label{MIMO_Asym}
\end{minipage}
\end{figure}

Fig. \ref{MIMO_Symm} shows the average SER performance for the two AS schemes in the MIMO-PNC setting ($N_{R} > 1$) for symmetric channels. It is clear from the figure that the average SER for the MIMO-PNC system with AS1 decays as $(E_{\min}/N_{0})^{-2}$ for higher values of $E_{\min}/N_{0}$ when $(N_{A}, N_{R}, N_{B})$ is $(2,2,2), \,(2, 2, 3)$ or $(3, 2, 3)$, and it is clear that the system fails to achieve transmit diversity -- the diversity order of the system depends only on the number of antennas at the relay. On the other hand, for the MIMO-PNC system with AS2, the average SER decays as $(E_{\min}/N_{0})^{-4}$ for higher values of $E_{\min}/N_{0}$ when $(N_{A}, N_{R}, N_{B})$ is $(2, 2, 2)$ or $(2, 2, 3)$. Similarly, the average SER decays as $(E_{\min}/N_{0})^{-6}$ for higher values of $E_{\min}/N_{0}$ when $(N_{A}, N_{R}, N_{B})$ is $(3, 2, 3)$. Therefore, it is clear from Fig. \ref{MIMO_Symm}, that in case of MIMO-PNC, the diversity order is equal to $N_{R}$ with AS1 and $\min(N_{A}, N_{B}) \times N_{R}$ with AS2 resulting in the performance superiority of AS2.

Fig. \ref{MIMO_Asym} shows the average SER performance comparison of the two AS schemes in asymmetric channels with different number of user antennas in the MIMO-PNC setting (i.e., $N_{R} > 1$). In this case, $E_{A} / N_{0} = (E_{\min}/N_{0}) + 5$ dB (and thus $E_{B}/N_{0} = E_{\min}/N_{0}$). Similar to the previous results for the MIMO-PNC system in symmetric channels, the diversity order achieved by the MIMO-PNC system with AS1 for asymmetric channels is 2 ($= N_{R}$) for the case when $(N_{A}, N_{R}, N_{B})$ is $(2, 2, 2)$, $(3, 2, 2)$ or $(3, 2, 3)$. On the other hand, the diversity order achieved by the MIMO-PNC system with AS2 is 4 for the case when $(N_{A}, N_{R}, N_{B})$ is $(2, 2, 2)$ or $(3, 2, 2)$, while the system achieves a diversity order of 6 when $(N_{A}, N_{R}, N_{B})$ is $(3, 2, 3)$. Hence, it is clear that the MIMO-PNC system with AS2 achieves full diversity order of $\min(N_{A}, N_{B}) \times N_{R}$ and therefore outperforms the MIMO-PNC system with AS1 in asymmetric channels also.
\section{Conclusion}
In this paper, we analyzed the error performance of a MIMO-PNC system with fixed network coding under two different user-antenna selection schemes in an asymmetric scenario, where the users may have different number of antennas and different average SNR to the relay, this analysis being valid for any modulation order $M$. A detailed investigation of the error performance and the diversity order was presented. It was shown analytically that for the first antenna selection scheme (AS1), where each user-antenna is selected to maximize the overall channel gain between the user and the relay, the MIMO-PNC system achieves full diversity order of $\min(N_{A}, N_{B}) \times N_{R}$ for binary modulations. For non-binary modulations, a closed-form expression for a tight upper bound on the average SER was derived for the special case $N_{R} = 1$, and diversity analysis confirmed that the system fails to achieve transmit diversity. Simulation results show that the derived upper bound is tight and that for $N_R > 1$ the system fails to achieve transmit diversity, as the diversity order of the system remains equal to $N_R$. To overcome the aforementioned problems, a Euclidean distance based user-antenna selection scheme (AS2) was proposed which outperforms the first scheme in terms of error performance. Upper and lower bounds on the resulting average SER were derived, and it was shown that the MIMO-PNC system with AS2 achieves both transmit and receive diversity, resulting in a full diversity order of $\min(N_{A}, N_{B}) \times N_{R}$. This new user-antenna selection scheme allows the MIMO-PNC system to avoid the harmful effect of singular fade states without any need for adaptive network codes or nonstandard constellation design, reducing the overall system complexity.

%
%
\appendices
\section{Proof of Proposition 1}
Since $\mathfrak{h}_{m, i, j} \sim \mathcal{CN}(0, 1)$, the magnitude of $\mathfrak{h}_{m, i, j}$ is Rayleigh distributed. Defining $\omega_{m, i} \triangleq \sum_{j = 1}^{N_{R}} |\mathfrak{h}_{m, i, j}|^{2}$, then $\omega_{m, i} \sim Ga(N_{R}, 1)$. Therefore the PDF\footnote{We use a more accurate probabilistic notation here: for a random variable $\Lambda$, we denote the CDF by $F_{\Lambda}(\tau) = \mathrm{Pr}(\Lambda \leq \tau)$ and the PDF by $f_{\Lambda}(\tau) = \dfrac{d}{d\tau}F_{\Lambda}(\tau)$.} of $\omega_{m, i}$ can be written as, 
\begin{equation}
	f_{\omega_{m, i}}(\tau) = \dfrac{1}{\Gamma(N_{R})} \tau^{N_{R} - 1} \exp(-\tau). \notag
\end{equation}
For $N_{R} > 0$, the cumulative distribution function (CDF) is a special case of that of an Erlang distribution, i.e.,
\begin{equation}
	F_{\omega_{m, i}}(\tau) = 1 - \exp (-\tau) \sum_{j = 0}^{N_{R} - 1}\dfrac{\tau^{j}}{j!}. \notag
\end{equation}
Using \eqref{z_def}, $\mathfrak{z}_{m} = \sum_{j = 1}^{N_{R}} |h_{m, j}|^{2}$, we have 
\begin{align}
	& f_{\mathfrak{z}_{m}}(\tau) = \dfrac{d}{d\tau} \prod_{i = 1}^{N_{m}} F_{\omega_{m, i}}(\tau) = \dfrac{d}{d\tau} \left[ 1 - \exp (-\tau) \sum_{j = 0}^{N_{R} - 1}\dfrac{\tau^{j}}{j!} \right]^{N_{m}} \notag \\
	= & N_{A} \underbrace{\left[\! 1\! -\! \exp (-\tau)\!\! \sum_{j = 0}^{N_{R} - 1}\!\!\dfrac{\tau^{j}}{j!} \!\right]^{\!N_{m} - 1}}_{\mathfrak{X}_{1}}  \underbrace{\dfrac{d}{d \tau}\!\! \left[\! 1\! -\! \exp (-\tau)\!\! \sum_{j = 0}^{N_{R} - 1}\!\!\dfrac{\tau^{j}}{j!} \right]}_{\mathfrak{X}_{2}}. \label{pdf_z_tau}
\end{align}
Using the multinomial theorem, 
\begin{align}
	\mathfrak{X}_{1} & = \sum_{ \substack{ k_{0}+ k_{1} + \cdots + k_{N_{R}} \\ = N_{m} - 1} } \binom{N_{m} - 1}{k_{0}, \ldots, k_{N_{R}}} (-1)^{N_{m} - 1 - k_{0}}  \left[ \prod_{j = 0}^{N_{R} - 1} \left( \dfrac{1}{j!}\right)^{k_{j + 1}} \right]  \tau^{s} \exp \left[ -(N_{m} - 1 - k_{0}) \tau\right], \label{X1}
\end{align}
where $s = \sum_{j = 0}^{N_{R}}j k_{j + 1}$. Moreover, 
\begin{equation}
	\mathfrak{X}_{2} = \dfrac{1}{(N_{R} - 1)!} \tau^{N_{R} - 1} \exp(-\tau). \label{X2}
\end{equation}
Using \eqref{pdf_z_tau}, \eqref{X1} and \eqref{X2}, the closed-form expression for $f(\mathfrak{z}_{m})$ becomes equal to \eqref{fz}.
\section{Derivation of the Closed-Form Expression for $\Theta_{A, 1}$}
For $\mathfrak{z}_{A} = \sum_{j = 1}^{N_{R}} \left\vert h_{A, j}\right\vert^{2}$, \eqref{Upsilon1_Case1} can be rewritten as
\begin{align}
	& \Upsilon_{1} = \exp \left( - \dfrac{E_{A} |\Delta x_{A}|^{2}}{4N_{0}} \mathfrak{z}_{A}\right) \notag \\
	\implies & \mathbb{E}[\Upsilon_{1}] = \int_{0}^{\infty} \exp \left( - \dfrac{E_{A} |\Delta x_{A}|^{2}}{4N_{0}} \mathfrak{z}_{A}\right) f(\mathfrak{z}_{A}) \, d\mathfrak{z}_{A}. \label{Avg_Upsilon1}
\end{align}
Substituting the expression \eqref{fz} for $f(\mathfrak{z}_{A})$ into \eqref{Avg_Upsilon1} yields
\begin{align}
	\mathbb{E}[\Upsilon_{1}] = & \dfrac{N_{A}}{(N_{R} - 1)!} \sum_{\substack{ k_{0} + k_{1} + \cdots + k_{N_{R}} \\ = N_{A} - 1}} \binom{N_{A} - 1}{k_{0}, \ldots, k_{N_{R}}} (-1)^{N_{A} - 1- k_{0}}  \left[ \prod_{j = 0}^{N_{R} - 1} \left( \dfrac{1}{j!}\right)^{k_{j + 1}}\right]  \notag \\
	& \hspace{5cm} \times \int_{0}^{\infty} \mathfrak{z}_{A}^{N_{R} + s - 1} \exp \left[- \left(\dfrac{E_{A} |\Delta x_{A}|^{2}}{4N_{0}} + N_{A} - k_{0} \right)\mathfrak{z}_{A} \right]\, d\mathfrak{z}_{A} \notag \\
	= & \dfrac{N_{A}}{(N_{R} - 1)!} \sum_{\substack{ k_{0} + k_{1} + \cdots + k_{N_{R}} \\ = N_{A} - 1}} \binom{N_{A} - 1}{k_{0}, \ldots, k_{N_{R}}} (-1)^{N_{A} - 1- k_{0}} \left[ \prod_{j = 0}^{N_{R} - 1} \left( \dfrac{1}{j!}\right)^{k_{j + 1}}\right] \notag \\
	& \hspace{6cm} \times \left(\dfrac{E_{A} |\Delta x_{A}|^{2}}{4N_{0}} + N_{A} - k_{0} \right)^{-(N_{R} + s)} (N_{R} + s - 1)!. \notag
\end{align} 
The integration above is solved using~\cite[p.~322]{Prudnikov} to obtain~\eqref{ThetaA1}.

\ifCLASSOPTIONcaptionsoff
  \newpage
\fi
\bibliographystyle{IEEEtran}
\balance
\bibliography{PNC_TCOMM}

\begin{thebibliography}{10}
\providecommand{\url}[1]{#1}
\csname url@samestyle\endcsname
\providecommand{\newblock}{\relax}
\providecommand{\bibinfo}[2]{#2}
\providecommand{\BIBentrySTDinterwordspacing}{\spaceskip=0pt\relax}
\providecommand{\BIBentryALTinterwordstretchfactor}{4}
\providecommand{\BIBentryALTinterwordspacing}{\spaceskip=\fontdimen2\font plus
\BIBentryALTinterwordstretchfactor\fontdimen3\font minus
  \fontdimen4\font\relax}
\providecommand{\BIBforeignlanguage}[2]{{%
\expandafter\ifx\csname l@#1\endcsname\relax
\typeout{** WARNING: IEEEtran.bst: No hyphenation pattern has been}%
\typeout{** loaded for the language `#1'. Using the pattern for}%
\typeout{** the default language instead.}%
\else
\language=\csname l@#1\endcsname
\fi
#2}}
\providecommand{\BIBdecl}{\relax}
\BIBdecl

\bibitem{Liew}
S.~C. Liew, S.~Zhang, and L.~Lu, ``Physical-layer network coding: Tutorial,
  survey, and beyond,'' \emph{Physical Communication}, vol.~6, pp. 4 -- 42,
  2013.

\bibitem{Popovski}
P.~Popovski and H.~Yomo, ``The anti-packets can increase the achievable
  throughput of a wireless multi-hop network,'' in \emph{Proc. IEEE Int. Conf.
  Communications (ICC)}, vol.~9, Jun. 2006, pp. 3885--3890.

\bibitem{Primer}
S.~C. Liew, Lu, and S.~Zhang, \emph{A Primer on Physical-Layer Network
  Coding}.\hskip 1em plus 0.5em minus 0.4em\relax Morgan \& Claypool, 2015.

\bibitem{FieldPNC}
S.~Zhang, S.~C. Liew, and L.~Lu, ``Physical layer network coding schemes over
  finite and infinite fields,'' in \emph{Proc. IEEE Global Commun. Conf.
  (GLOBECOM)}, Nov. 2008, pp. 1--6.

\bibitem{Katti}
S.~Katti, S.~Gollakota, and D.~Katabi, ``Embracing wireless interference:
  Analog network coding,'' \emph{ACM SIGCOMM Comput. Commun. Rev.}, vol.~37,
  no.~4, pp. 397--408, Aug. 2007.

\bibitem{Branka_Globecom}
R.~H.~Y. Louie, Y.~Li, and B.~Vucetic, ``Performance analysis of physical layer
  network coding in two-way relay channels,'' in \emph{Proc. IEEE Global
  Commun. Conf. (GLOBECOM)}, Nov. 2009, pp. 1--6.

\bibitem{MinKim}
M.~Ju and I.~M. Kim, ``Error performance analysis of {BPSK} modulation in
  physical-layer network-coded bidirectional relay networks,'' \emph{IEEE
  Trans. Commun.}, vol.~58, no.~10, pp. 2770--2775, Oct. 2010.

\bibitem{PNC_BER}
M.~Park, I.~Choi, and I.~Lee, ``Exact {BER} analysis of physical layer network
  coding for two-way relay channels,'' in \emph{Proc. IEEE Veh. Technol. Conf.
  (VTC Spring)}, May 2011, pp. 1--5.

\bibitem{SER}
K.~Lu, S.~Fu, Y.~Qian, and H.~H. Chen, ``{SER} performance analysis for
  physical layer network coding over {AWGN} channels,'' in \emph{Proc. IEEE
  Global Commun. Conf. (GLOBECOM)}, Nov. 2009, pp. 1--6.

\bibitem{Andrew}
K.~Ravindran, A.~Thangaraj, and S.~Bhashyam, ``High {SNR} error analysis for
  bidirectional relaying with physical layer network coding,'' \emph{IEEE
  Trans. Commun.}, vol.~65, no.~4, pp. 1536--1548, Apr. 2017.

\bibitem{LinearVectorPNC}
J.~Guo, T.~Yang, J.~Yuan, and J.~A. Zhang, ``Linear vector physical-layer
  network coding for {MIMO} two-way relay channels: Design and performance
  analysis,'' \emph{IEEE Trans. Commun.}, vol.~63, no.~7, pp. 2591--2604, July
  2015.

\bibitem{MIMO_LinearPNC}
L.~Shi, T.~Yang, K.~Cai, P.~Chen, and T.~Guo, ``On {MIMO} linear physical-layer
  network coding: Full-rate full-diversity design and optimization,''
  \emph{IEEE Trans. Wireless Commun.}, vol.~17, no.~5, pp. 3498--3511, May
  2018.

\bibitem{RajanPNC}
V.~T. Muralidharan and B.~S. Rajan, ``Performance analysis of adaptive physical
  layer network coding for wireless two-way relaying,'' \emph{IEEE Trans.
  Wireless Commun.}, vol.~12, no.~3, pp. 1328--1339, Mar. 2013.

\bibitem{Akino}
T.~Koike-Akino, P.~Popovski, and V.~Tarokh, ``Optimized constellations for
  two-way wireless relaying with physical network coding,'' \emph{IEEE J. Sel.
  Areas Commun.}, vol.~27, no.~5, pp. 773--787, Jun. 2009.

\bibitem{AdaptiveOFDM}
H.~Yan and H.~H. Nguyen, ``Adaptive physical-layer network coding in two-way
  relaying with {OFDM},'' in \emph{Proc. IEEE Global Comm. Conf. (GLOBECOM)},
  Dec. 2013, pp. 4244--4249.

\bibitem{RajanMIMO}
V.~T. Muralidharan and B.~S. Rajan, ``Wireless network coding for {MIMO}
  two-way relaying,'' \emph{IEEE Trans. Wireless Commun.}, vol.~12, no.~7, pp.
  3566--3577, Jul. 2013.

\bibitem{HePNC_Rajan}
C.~Arunachala, S.~D. Buch, and S.~Rajan, ``Wireless bidirectional relaying
  using physical layer network coding with heterogeneous {PSK} modulation,''
  \emph{IEEE Trans. Veh. Technol.}, to appear.

\bibitem{MultiWay}
V.~T. Muralidharan and B.~S. Rajan, ``Physical layer network coding for
  wireless multi-way relaying with finite signal sets,'' \emph{Int. J. Advances
  in Engineering Sciences and Applied Mathematics}, vol.~5, pp. 2--11, Mar.
  2013.

\bibitem{LatinSquares}
V.~T. Muralidharan, V.~Namboodiri, and B.~S. Rajan, ``Wireless network-coded
  bidirectional relaying using latin squares for $m$-{PSK} modulation,''
  \emph{IEEE Trans. on Inf. Theory}, vol.~59, no.~10, pp. 6683--6711, Oct.
  2013.

\bibitem{Huang}
M.~Huang and J.~Yuan, ``Error performance of physical-layer network coding in
  multiple-antenna {TWRC},'' \emph{IEEE Trans. Veh. Technol.}, vol.~63, no.~8,
  pp. 3750--3761, Oct. 2014.

\bibitem{Paulraj}
R.~W. Heath and A.~Paulraj, ``Antenna selection for spatial multiplexing
  systems based on minimum error rate,'' in \emph{Proc. IEEE Int. Conf. Commun.
  (ICC)}, vol.~7, 2001, pp. 2276--2280.

\bibitem{PNC_Scheduling}
Y.~Jeon, Y.~T. Kim, M.~Park, and I.~Lee, ``Opportunistic scheduling for
  three-way relay systems with physical layer network coding,'' in \emph{Proc.
  IEEE Veh. Technol. Conf. (VTC Spring)}, May 2011, pp. 1--5.

\bibitem{Hari}
R.~Rajashekar, K.~V.~S. Hari, and L.~Hanzo, ``Quantifying the transmit
  diversity order of {E}uclidean distance based antenna selection in spatial
  modulation,'' \emph{IEEE Signal Process. Lett.}, vol.~22, no.~9, pp.
  1434--1437, Sept. 2015.

\bibitem{Chiani}
M.~Chiani, D.~Dardari, and M.~K. Simon, ``New exponential bounds and
  approximations for the computation of error probability in fading channels,''
  \emph{IEEE Trans. Wireless Commun.}, vol.~2, no.~4, pp. 840--845, Jul. 2003.

\bibitem{Stegun}
M.~Abramowitz and I.~A. Stegun, \emph{Handbook of Mathematical Functions with
  Formulas, Graphs, and Mathematical Tables}, 9th~ed.\hskip 1em plus 0.5em
  minus 0.4em\relax New York: Dover, 1964.

\bibitem{PrudnikovVol2}
A.~Prudnikov and O.~Marichev, \emph{Integrals and Series: Volume 2: Special
  functions}.\hskip 1em plus 0.5em minus 0.4em\relax Gordon and Breach Science
  Publishers, 1998.

\bibitem{Giannakis}
Z.~Wang and G.~B. Giannakis, ``A simple and general parameterization
  quantifying performance in fading channels,'' \emph{IEEE Trans. Commun.},
  vol.~51, no.~8, pp. 1389--1398, Aug. 2003.

\bibitem{UpBound}
X.~Jin and D.~Cho, ``Diversity analysis on transmit antenna selection for
  spatial multiplexing systems with {ML} detection,'' \emph{IEEE Trans. Veh.
  Technol.}, vol.~62, no.~9, pp. 4653--4658, Nov. 2013.

\bibitem{Prudnikov}
A.~Prudnikov, \emph{Integrals and Series: Volume 1: Elementary
  Functions}.\hskip 1em plus 0.5em minus 0.4em\relax Taylor \& Francis, 1986.

\end{thebibliography}

\end{document}